\documentclass[a4paper,12pt]{article}
\usepackage{amsfonts}
\usepackage{amssymb}
\usepackage{amsmath}
\usepackage{amsthm}
\usepackage{enumerate}
\usepackage{graphicx}
\usepackage{amsmath}
\usepackage{amssymb}
\usepackage{amsfonts}
\usepackage{amsthm}
\usepackage{epsfig}
\usepackage{graphicx}

\newtheorem{teor}{Theorem}[section]

\newtheorem{rem}{Remark}[section]
\newtheorem*{col*}{Collorary}
\newcommand{\ee}{\mathrm{e}}
\newcommand{\ii}{\mathrm{i}}
\renewcommand{\Im}{\operatorname{Im}}

 \providecommand{\Db}{\mathbb}

\newcommand{\D}{\displaystyle}
\providecommand{\DS}{\displaystyle}
\newcommand{\ord}{\mathrm{O}}
\renewcommand{\Re}{\operatorname{Re}}
\begin{document}
\title{Shock problem for MKdV equation: Long-time Dynamics of the Step-like Initial Data}
\author{Vladimir Kotlyarov and Alexander Minakov
\\[1mm]
{\scriptsize Mathematical division, Institute for Low Temperature
Physics}
\\[-2mm]
{\scriptsize 47 Lenin Avenue, 61103 Kharkiv, Ukraine}} \maketitle
\begin{abstract}
We consider the modified Korteveg de Vriez equation on the whole
line. Initial data is real and step-like, i.e. $q(x,0)=0$ for
$x\geq0$ and $q(x,0)=c$ for $x<0$, where c is arbitrary real
number. The goal of this paper is to study the asymptotic
behavior of the initial-value problem's solution by means of the
asymptotic behavior of the some Riemann\textendash Hilbert
problem. In this paper we show that the solution of this problem
has different asymptotic behavior in different regions. In the
region $x<-6c^2t$ and $x>4c^2t$ the solution is tend to $c$ and
$0$ correspondingly. In the region $-6c^2t<x<4c^2t$ the solution
takes the form of a modulated elliptic wave.
\end{abstract}

\section{Introduction}
\setcounter{equation}{0}

Initial value problems with step-like initial function have very
long story beginning from the papers by A.V. Gurevich, L.P.
Pitaevsky \cite{GP} and E.Ya.Khruslov \cite{Kh} in a middle of
70th. There are many papers devoted to different aspects of these
problems. For the present time there are a few full and rigorous
results on an asymptotic behavior of such problems \cite{KK1},
\cite{Bik1}-\cite{Bik5}, \cite{BV},.....,. We pay our attention
here to a problem which was not considered elsewhere and give
results in a rigorous form using the method of the matrix
Riemann-Hilbert problem and the steepest descent method for
oscillatory matrix RH problems. We consider the modified Korteweg
de Vries equation on the whole line. Initial datum is a
step-like, i.e. $q(x,0)=0$ for $x\ge0$ and $q(x,0)= c$ for $x<0$,
where $c$ is an arbitrary real number. Without loss of generality
we put $c>0$. This problem can be considered as a shock problem.
The goal of this paper is to study asymptotic behavior of the
Riemann \textendash Hilbert problem whose solution gives the
solution of the initial-value problem. In this paper we show that
the solution of the shock problem has different asymptotic
behavior in different regions. In the regions $-\infty<x<-6c^2t$
and $4c^2t<x<\infty$, the solution is trivial, i.e. it is equal
to $c$ and $0$ respectively. In the region $-6c^2t<x<-4c^2t$ the
solution takes the form of a modulated elliptic wave of finite
amplitude. Thus for a large time the solution has finite
amplitude in the first two regions while in the third region
($4c^2t<x<\infty$) it takes the form of a  vanishing (as
$\ord(t^{-1/2})$) self-similar wave. The development of the
Riemann-Hilbert method for the shock problems arising  for
integrable PDEs on the whole line with the different finite-gap
boundary conditions as $x \to \pm \infty$ goes back to the works
done in 80-90s by R. Bikbaev, P. Deift, V. Novokshenov, and S.
Venakides. All those results was devoted to the initial-value
problem with self-adjoint Lax operators. Most recently, an
implementation of the RH scheme to the shock problem and the
evaluation of the long-time asymptotics of the solution to the
focusing nonlinear Schr\"odinger equation on the whole line,
where the Lax operator is not self-adjoint and the initial
function was chosen in the form $q(x,0)=A\exp{(i\mu\vert
x\vert)}$, was done in \cite{BV}. It is worth mentioning that our
shock problem is different from that in \cite{BV} as well as our
construction of the phase  $g$ - function.

The inverse scattering transform method (IST) for solving
initial-value   problems for nonlinear evolutionary equations,
discovered in 1967 \cite{GGKM}, turned out to be a very powerful
tool, which allowed to obtain a huge number of very interesting
results in different areas of mathematics and physics. At the
beginnig of 90th a new great achievement in the further
development of the IST method have been done by P.Deift and
X.Zhou. It is a nonlinear steepest descent method for oscillatory
matrix Riemann-Hilbert problem. With the new method it came a
nice possibility to rewrite known asymptotic results for
different nonlinear integrable models in the rigorous and
transparent form (sf.\cite{DZ92},\cite{DZ93},\cite{DIZ93}) and
obtain numerous new significant results in the theory of
completely integrable nonlinear equations, random matrix models
and orthogonal polynomials, integrable statistical mechanics. Our
goal is to bring new results in theory of the shock problems,
especially in the case of non self-adjoint Lax operators, and
some development of ideas, given recently in \cite{BIK},
\cite{BIK1} in the direction of the strengthening  of the
nonlinear steepest descent method for oscillatory matrix
Riemann-Hilbert problem.

 Let us consider the problem
\begin{equation}\label{mkdv}
q_t+6q^2q+q_{xxx}=0
\end{equation}
\begin{equation}\label{iv}
q(x,0)=q_0(x)\rightarrow\begin{cases} 0,\quad x\rightarrow+\infty \\
c,\quad x \rightarrow-\infty,\end{cases}
\end{equation}
where $q_0(x)$ is arbitrary step-like function, including the
discontinuous case: $q_0(x)\equiv c$ for $x<0$ and
$q_0(x)\equiv0$ for $x\ge0$. We suppose that the solution
$q(x,t)$ of this problem exists for $x \in \mathbb{R},t
\in\mathbb{R}_+$. To study  the initial value problem
(\ref{mkdv})-(\ref{iv}) we will use the Lax representation of the
MKdV equation in the form of over-determined system of
differential equations:
\begin{equation}\label{xeq}\Phi_x +
ik\sigma_3\Phi  =Q(x,t)\Phi
\end{equation}
\begin{equation}\label{teq}\Phi_t +4ik^3\sigma_3\Phi
=\hat Q(x,t,k)\Phi ,\end{equation} where $\Phi =\Phi(x,t,k)$ is a
$2\times 2$ matrix-valued function,
\begin{equation}\label{Q}
\sigma_3:=
\begin{pmatrix}1 &0\\0& -1\end{pmatrix}, \qquad
Q(x,t):=\begin{pmatrix} 0 & q(x,t) \\
-q(x,t) & 0\end{pmatrix},
\end{equation}
\begin{equation}\label{Qhat}\hat Q(x,t,k)=4k^2Q(x,t,k)-
2ik(Q^2(x,t,k)+Q_x(x,t,k))\sigma_3+ 2Q^3(x,t,k)-Q_{xx}(x,t,k),
\end{equation}
and $k \in \mathbb{C}$. The equations (\ref{xeq}) and (\ref{teq})
are compatible if and only if the function $q(x,t)$ satisfy the
MKdV equation (\ref{mkdv}). To apply the inverse scattering
transform to the problem (\ref{mkdv}), (\ref{iv}) we have to
construct matrix valued solution of this equations defined by
their asymptotics:
\begin{equation}\label{Phi}
\Phi(x,t,k)=e^{-(ikx+4ik^3t)\sigma_3}+\mathrm{O}
\left(\frac{1}{k}\right),\quad x\to+\infty,\quad \Im k=0
\end{equation}
\begin{equation}\label{Psi}
\Psi(x,t,k)=E(x,t,k)+ {\mathrm{ O}}\left(\frac{1}{k}\right),
\quad x\rightarrow-\infty,\quad \Im k=0.
\end{equation}
Here $E(x,t,k)$ is the solution of the   linear differential
equations
\begin{equation}\label{xeqc}\\E_x +
ik\sigma_3E =Q_c E
\end{equation}
\begin{equation}\label{teqc}E_t +4ik^3\sigma_3E
=\hat Q_c(k) E ,\end{equation} where
\begin{equation}\label{Q_c}
Q_c :=\begin{pmatrix} 0 & c \\
-c & 0\end{pmatrix}\end{equation}
\begin{equation}\label{Q_chat}\hat Q_c( k)=4k^2Q_c-2ikQ_c^2
\sigma_3+2Q_c^3.
\end{equation}
We chose the solution $E(x,t,k)$ as follows:
\begin{equation}\label{E}E(x,t,k)=\displaystyle\frac{1}{2}
\left(\begin{array}{ccc}\varkappa(k)+\displaystyle\frac{1}
{\varkappa(k)}&
\varkappa(k)-\displaystyle\frac{1}{\varkappa(k)}\\\\
\varkappa(k)-\displaystyle\frac{1}{\varkappa(k)}&
\varkappa(k)+\displaystyle\frac{1}{\varkappa(k)}
\end{array}\right)
e^{-ixX(k)\sigma_3-it\Omega(k)\sigma_3},\end{equation} where
\begin{equation}\label{XOk}X(k)=\sqrt{k^2+c^2},\qquad \Omega(k)=
2(2k^2-c^2)X(k),
\qquad\varkappa(k)=\sqrt[4]{\displaystyle\frac{k-ic}{k+ic}}
\end{equation}
$X(k)$ and $\varkappa(k)$ are analytic in the complex plane cut
along the segment $[ic,-ic]$, i.e. $ k\in\mathbb{C}
\setminus[-ic,ic]$, and branches of roots are as follows:
$X(1)>0$, $\varkappa(\infty)=1$.

The solution $\Phi(x,t,k)$ can be represented in the form:
\begin{equation}
\label{f1} \Phi(x,t,k)=\Big(\ee^{-\ii kx\sigma_{3}}+
\int\limits_{x}^{\infty}K(x,y,t)\ee^{-\ii ky\sigma_{3}}dy\Big)
\ee^{-4\ii k^3t\sigma_{3}},
\end{equation}
where the kernel $K(x,y,t)$ is chosen to be so that the first
factor satisfies the $x-$equation (\ref{xeq}) for all $t$, and
the second factor satisfies the $t-$equation (\ref{teq}) for
$x=\infty$. Then, by the same way as in  \cite{BMK}, we prove
that $\Phi(x,t,k)$ satisfies both equations (\ref{xeq}) and
(\ref{teq}). The existence of the solution represented by the
transformation operators with the kernel $K(x,y,t)$ is proved  in
\cite{BMK}.  By the same manner another solution takes the form:
\begin{equation}
\label{f1} \Psi(x,t,k)=  E(x,t,k)+
\int\limits_{-\infty}^{x}L(x,y,t)E(y,t,k)dy
\end{equation}
with some matrix kernel $L(x,y,t)$. Omitting rutin details of the
proof of these representations we formulate below properties of
the solutions.

The matrices $\Phi(x,t,k)$ and $\Psi(x,t,k)$  defined by
(\ref{Phi}) and (\ref{Psi}) and their columns $\Phi_j(x,t,k)$ and
$\Psi_j(x,t,k)$, $j=1,2$ have the following properties:
\begin{enumerate}\label{propPhiPsi}
  \setcounter{enumi}{0}

  \item determinants:\\
  $\det\Phi(x,t,k)=1,\quad \det\Psi(x,t,k)=1$.

  \item analyticity:\\
  $\Phi_1(x,t,k)$ is analytic in $k\in\Db{C}_-$,$\quad
  \Phi_2(x,t,k)$ is analytic in $k\in\Db{C}_+$,\\
  $\quad \Psi_1(x,t,k)$ is analytic in  $k\in\Db{C}_+\setminus [0,ic]$,
  $\quad \Psi_2(x,t,k)$ is analytic in $k\in\Db{C}_-\setminus [-ic,0]$,\\
  $\quad \Psi_1(x,t,.)$ and $\Psi_2(x,t,.)$ have continuous  extensions
  to $(-ic,ic)_-\cup(-ic,ic)_+$.

  \item symmetries:
  \[\begin{array}{llll}
  \overline{\Psi_{22}(x,t,\overline{k})}&=\Psi_{11}(x,t,k),
       &\Psi_{22}(x,t,-k)&=\Psi_{11}(x,t,k),\\
      \overline{\Psi_{12}(x,t,\overline{k})}&=-\Psi_{21}(x,t,k),
      &\Psi_{12}(x,t,-k)&=-\Psi_{21}(x,t,k),\\
    \overline{\Phi_{22}(x,t,\overline{k})}&=\Phi_{11}(x,t,k),
      &\Phi_{22}(x,t,-k)&=\Phi_{11}(x,t,k),\\
      \overline{\Phi_{12}(x,t,\overline{k})}&=-\Phi_{21}(x,t,k),
      &\Phi_{12}(x,t,-k)&=-\Phi_{21}(x,t,k),\\
      \overline{\Phi_{jl}(x,t,-\overline{k})}&=\Phi_{jl}(x,t,k),&\quad
      j,l=\overline{1,2}.
\end{array}\]
  \item large $k$ asymptotics:
  \[
  \left.\begin{matrix}\Phi_1(x,t,k)e^{+ikx+4ik^3t}&\\
  \Psi_2(x,t,k)e^{-ikx-4ik^3t}&\end{matrix}\right\}
  =1+{\mathrm{O}}\left(\displaystyle\frac{1}{k}
      \right),\quad k\rightarrow\infty, \quad \Im k\le0;
 \]
 \[
  \left.\begin{matrix}\Psi_1(x,t,k)e^{+ikx+4ik^3t}&\\
  \Phi_2(x,t,k)e^{-ikx-4ik^3t}&\end{matrix}\right\}
  =1+{\mathrm{O}}\left(\displaystyle\frac{1}{k}
      \right),\quad k\rightarrow\infty, \quad \Im k\ge0;
 \]
  \item jump: \\
  $\Psi_-(x,t,k)=\Psi_+(x,t,k)\left(
  \begin{array}{ccc}0&i\\i&0\end{array}\right),\quad
  k\in(-ic,ic)$,
  where $\Psi_\pm(x,t,k)$ are the boundary values of the matrix
  $\Psi(x,t,k)$ from the left ($+$) and from the right ($-$) of
  the  oriented downwards  interval $(-ic, ic)$.
\end{enumerate}

The matrices $\Phi(x,t,k)$ and $\Psi(x,t,k)$ are the solution of
equations (\ref{xeq}) and (\ref{teq}). Hence they are linear
dependent, i.e. there exists the independent on $x,t$ matrix:
\begin{equation}\label{T}T(k)=\Phi^{-1}(x,t,k)\Psi(x,t,k),\quad
k\in\mathbb{R}\end{equation} which is defined for real $k$. Some
of elements of this matrix have a larger domain of the
definition. Indeed, using  (\ref{T}) we find
\begin{equation}T_{11}(k)=\det(\Psi_1,\Phi_2)\end{equation}
\begin{equation}T_{21}(k)=\det(\Phi_1,\Psi_1)\end{equation}
\begin{equation}T_{12}(k)=\det(\Psi_2,\Phi_2)\end{equation}
\begin{equation}T_{22}(k)=\det(\Phi_1,\Psi_2).\end{equation}
Then the above properties of the solutions $\Phi(x,t,k)$ and
$\Psi(x,t,k)$ yield:
\begin{itemize}
       \item $T_{11}(k)$ is analytic in $k\in\Db{C}_+
       \backslash[0,ic]$ and has a continuous extension to $(0,ic)_-
       \bigcup(0,ic)_+$,

\item $T_{22}(k)$ is analytic in $k\in\Db{C}_- \backslash[0,ic]$
and has a continuous extension to $(-ic,0)_-\bigcup(-ic,0)_+$,

 \item $T_{21}(k)$ is continuous in
$k\in(-\infty,0)\bigcup(0,-ic]_-\bigcup[-ic,0)_+\bigcup(0,+\infty)$

\item $T_{11}(k)$ is continuous in
$k\in(-\infty,0)\bigcup(0,ic]_-\bigcup[ic,0)_+\bigcup(0,+\infty)$
\end{itemize}
and
\begin{itemize}
     \item $\overline{T_{22}(\overline{k})}=T_{11}(k)$,$\quad T_{22}(-k)=
     T_{11}(k),\\$
\item $\overline{T_{12}(\overline{k})}=-T_{21}(k)$,$\quad
T_{12}(-k)=-T_{21}(k),\\$ \item
$\overline{T_{jk}(-\overline{k})}=T_{jk}(k),\quad
j,k=\overline{1,2}$
\end{itemize}
Denote \[a(k)=T_{11}(k),\]\[b(k)=T_{21}(k).\] Define the
reflection coefficient
\[r(k)=\displaystyle\frac{b(k)}{a(k)}.\]
It has the following property:
\[\overline{r(-\overline{k})}=r(k)\]
The columns of the matrices $\Phi$ and $\Psi$ have the following
extra properties:
\begin{enumerate}\label{propPhiPsi}
\setcounter{enumi}{5}
  \item $\displaystyle\frac{(\Psi_1)_-(x,t,k)}{a_-(k)}-
      \displaystyle\frac{(\Psi_1)_+(x,t,k)}{a_+(k)}=f_1(k)\Phi_2(x,t,k),
      \quad k\in(0,ic)$
  \item $\displaystyle\frac{(\Psi_2)_-(x,t,k)}{\overline{a_-
  (\overline{k})}}-
      \displaystyle\frac{(\Psi_2)_+(x,t,k)}{\overline{a_+(\overline{k})}}=
      f_2(k)\Phi_2(x,t,k),\quad k\in(0,-ic)$
\end{enumerate}
where \[f_1(k)=\displaystyle\frac{i}{a_-(k)a_+(k)},\quad
k\in(0,ic)\]
\[f_2(k)=\displaystyle\frac{i}{\overline{a_-(\overline{k})}\overline{a_+
(\overline{k})}}=
-\overline{f_1(\overline{k})},\quad k\in(-ic,0)\]

\section{The Basic Riemann\textendash Hilbert problem}
\setcounter{equation}{0}

The scattering relations (\ref{T}) between matrix-valued
functions $\Psi(x,t,k)$ and $\Phi(x,t,k)$,  and also extra
properties 6, 7 can be rewritten in terms of the
Riemann\textendash Hilbert problem. To do so, let us define
matrix-valued function $M(\xi,t,k)$ by putting
\begin{equation}\label{M}M(\xi,t,k)=\begin{cases}\left(\displaystyle\frac{\Psi_1
(x,t,k)}{a(k)}
e^{it\theta(k,\xi)},\Phi_2(x,t,k)e^{-it\theta(k,\xi)}\right),\quad k\in \mathbb{C}_+ \backslash [0,ic]  \\\\
\left(\Phi_1(x,t,k)e^{it\theta(k,\xi)},\displaystyle\frac{\Psi_2(x,t,k)}
{\overline{a(\overline{k})}} e^{-it\theta(k,\xi)}\right),\quad
k\in\mathbb{C}_-\backslash[-ic,0], \end{cases}\end{equation}
where $x=12\xi t$ and $ \theta(k, \xi)=4k^3+12k\xi $
($\xi=x/12t$). To make the paper more transparent we consider
below only the  shock problem when the initial datum is
discontinuous:
\[q_0(x)=\begin{cases}0,\qquad x\geq0\\c,\qquad x<0.\end{cases}\]
Then
\begin{equation}\label{abr}
a(k)=\displaystyle\frac{1}{2}
\left(\varkappa(k)+\displaystyle\frac{1}{\varkappa(k)}\right),\qquad
 b(k)=\displaystyle\frac{1}{2}
\left(\varkappa(k)-\displaystyle\frac{1}{\varkappa(k)}\right),\qquad
r(k)=\displaystyle\frac{\varkappa^2(k)-1}{\varkappa^2(k)+1},
\end{equation}
where $\varkappa(k)$ is defined by (\ref{XOk}), are analytic in
$k\in\mathbb{C}\backslash[-ic,ic]$. The transition coefficient
$a^{-1}(k)$ is bounded in $k\in\mathbb{C}_+\backslash[ic,0]$
because the function $a(k)$ is equal to zero nowhere. In this
case we have:
\begin{equation}\label{fr}
f_2(k)=f_1(k)=r_-(k)-r_+(k),\quad k\in(-ic,ic)
\end{equation}

\begin{figure}[ht]
\begin{center}
\epsfig{width=100mm,figure=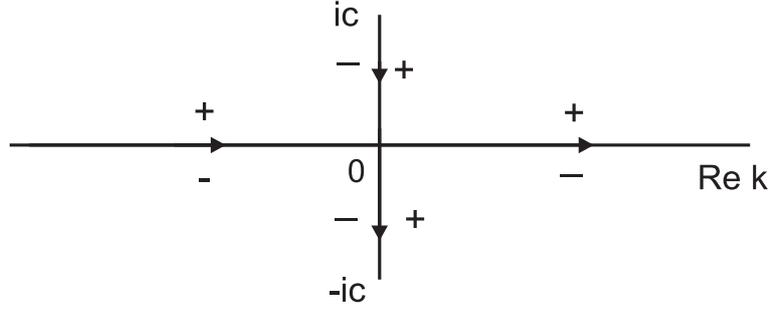}
\end{center}
\caption{The oriented contour $\Sigma$} \label{im-g1}
\end{figure}


Let us define the oriented contour $\Sigma=\Db{R}\cup[ic, -ic]$
as on the figure 1. Then the matrix $M (\xi,t,k)$  solves the
next Riemann\textendash Hilbert problem:
\begin{itemize}
\item matrix valued function $M (\xi,t,k)$ is analytic in the
domain $\mathbb{C}\setminus\Sigma$;

\item $M (\xi,t,k)$ is bounded in neighborhoods of the end points
$i c$ and $-ic$;

\item $M_{-}(\xi,t,k)=M_{+}(\xi,t,k)J (\xi,t,k), \quad
k\in\Sigma=\mathbb{R}\cup [ic, -ic] $, where\\
\\
\begin{equation}\label{RH}
J(\xi,t,k)=\begin{cases}\left(\begin{array}{ccc}1&-\overline{r(k)}
e^{-2it\theta(k,\xi)}\\\\-r(k)e^{2it\theta(k,\xi)}&1+|r(k)|^2\end{array}
\right),\quad
k\in\mathbb{R}\backslash \{0\}\\\\
\left(\begin{array}{ccc}1&0\\\\f(k)e^{2it\theta(k,\xi)}&1\end{array}\right),
\quad k\in[0,ic]\\\\
\left(\begin{array}{ccc}1&f
(k)e^{-2it\theta(k,\xi)}\\\\0&1\end{array}\right),\quad
k\in[0,-ic]
\end{cases}
\end{equation}
\item $M (\xi,t,k)=I+O(k^{-1}),\quad k\to\infty;$
\end{itemize}
where $r(k)={b(k)}/{a(k)}$, $k\in\mathbb{R}$ is given in
(\ref{abr}), and \[
f(k):=f_1(k)=f_2(k)=\displaystyle\frac{\ii}{a_-(k)a_+(k)}=
\displaystyle\frac{\ii}{\overline{a_-(\overline{k})}\
   \overline{a_+(\overline{k})}} \ ,\quad k\in[ic,-ic].
\]

If the initial datum is a generic step-like function then  $a(k)$
can have zeroes in the domain of analyticity. In this case the
matrix $M(\xi,t,k)$ will be meromorphic and residue relations
between columns of the matrix $M(\xi,t,k)$ must be added.

Now we forget about the origin of the Riemann-Hilbert problem and
suppose that the oriented contour $\Sigma$ and functions
(\ref{abr}) are given. The following theorem take place.

\begin{teor}
\label{thm:RH} Let the oriented contour $\Sigma$  and functions
(\ref{abr}) with
\[
\varkappa(k)=\sqrt[4]{\frac{k-ic}{k+ic}}
\]
 be given. Then the Riemann-Hilbert problem has a unique solution
$M(\xi,t,k)$.  The function $q(x,t)$, given by the equations,
\begin{equation}
\label{qM}
q(x,t)=2\ii\lim\limits_{k\to\infty}[kM(x/12t,t,k)]_{12},
\end{equation}
satisfy the MKdV equation (\ref{mkdv}) and the initial condition
\[
q(x,0)=q_0(x)=\begin{cases}0,\qquad x\geq0\\c,\qquad
x<0.\end{cases}
\]
\end{teor}

The proof of this theorem almost the same as in \cite{FT}, if we
takes into account that given functions $a(k)$, $b(k)$ via the
function $\varkappa(k)$ are in the one-to-one correspondence with
the shock function $q_0(x)$.

\section{Long time asymptotic analysis of the Riemann\textendash
Hilbert problem}

\subsection{ Steadiness region $x<-6c^2t$}

\begin{figure}[ht]
\begin{center}
\epsfig{width=100mm,figure=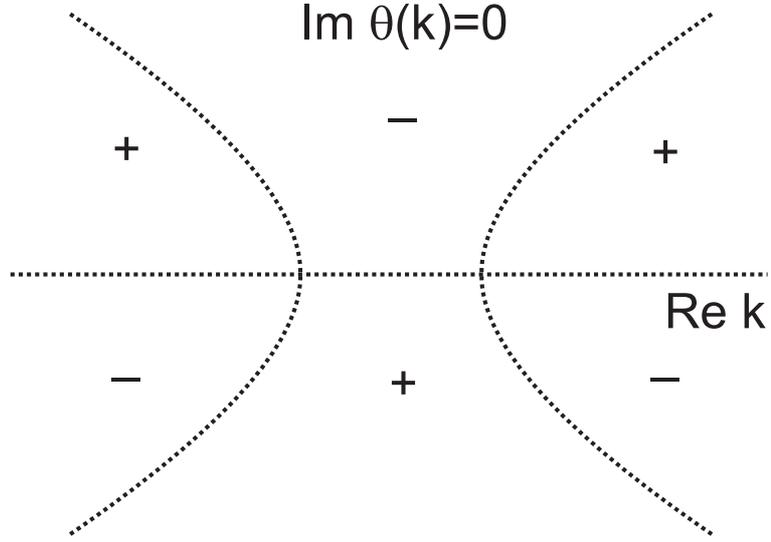}
\end{center}
\caption{The signature table of the function $\Im\theta(k)$}
\label{Graphic2}
\end{figure}

The jump matrices $J(\xi,t,k)$ depend on $\exp\{\pm2\ii
t\theta(k, \xi)\}$.  Hence the signature table of the imaginary
part of $ \theta(k,\xi)$  plays a very important role as the
phase function. The stationary points of the phase function
$\theta(k,\xi)$ equal to $\pm\sqrt{-\xi}$ and hence they are real
because $\xi<0$.  The signature table of the function
\[
\Im\theta(k,\xi)= (12\Re^2 k-4\Im^2 k+12\xi)\Im k
\]
depictured on the figure \ref{Graphic2}.

Thus $\Im\theta(k,\xi)>0$ ($\Im\theta(k,\xi)<0$) for $k$ lying in
the exterior (interior) of the hyperbola $3\Re^2 k-\Im^2
k+3\xi=0$ of the upper half-plane  and in the interior (exterior)
of the same hyperbola of the lower half-plane. For $\xi<0$
$\Im\theta(k,\xi)$ is negative along $(0,ic)$ and positive along
$(0,-ic)$. Therefore the jump matrices $J(\xi,t,k)$ are unbounded
(in $t$) when $k\in(-ic,ic)$. Hence we have to use the modified
nonlinear steepest descent method, suggested in \cite{DIZ},
\cite{DKMVZ}, \cite{BIK}, and find a new phase function $
g_c(k,\xi)$, instead of the function $\theta(k,\xi)$, which
transforms the original Riemann-Hilbert problem to the model RH
problem of the finite-gap type. New $g$-function leads to the
finite-gap model problems of zero genus for $\xi<\xi_-=-c^2/2 $
and genus one for $\xi_-<\xi<\xi_+=c^2/3 $. They are explicitly
solved by using elementary functions in the first region and the
elliptic theta functions in the second region, respectively.

In the  region $\xi < \xi_-$ we shall use the following
$g$-function:
\begin{equation}\label{gpw}
g_c(k,\xi)=\Omega(k)+ 12\xi X(k)=(4k^2-2c^2+12\xi)X(k)
\end{equation}
where $ X(k)=\sqrt{(k-ic )(k+ic)}=\sqrt{k^2+c^2}$. This function
has the asymptotic behavior similar to the phase function
$\theta(k,\xi)$, i.e.
\[
g(k,\xi)=4k^3+12\xi k + O(k^{-1}), \qquad k\to\infty.
\]
The differential of this function can be written in the form:
\[
dg_c(k,\xi)=\frac{12k^3+(6c^2+12\xi)k} {X(k)}dk=
\D\frac{(k+\lambda)k(k-\lambda)}{ X(k)}dk,
\]
where, evidently, $ \lambda(\xi) = \sqrt{-\xi-c^2/2}$. In order
to define the boundary value  $\xi_-$ we take into account that
the phase function $g_c(k,\xi)$ is acceptable until zeroes
$-\lambda(\xi)$ and $\lambda(\xi)$ are different. When they
coincide (are equal to zero ) and become complex conjugated then
$g_c(k,\xi)$ does not work and new phase function must be
introduced.  Hence we have to put $\lambda(\xi_-)= 0$. Then
$\xi_-=-c^2/2$, that gives $x<-6c^2t$, and the phase function
$g_c(k,\xi)$ will be useful for $\-\infty<\xi<\xi_-$.

\begin{figure}[ht]
\begin{center}
\epsfig{width=80mm,figure=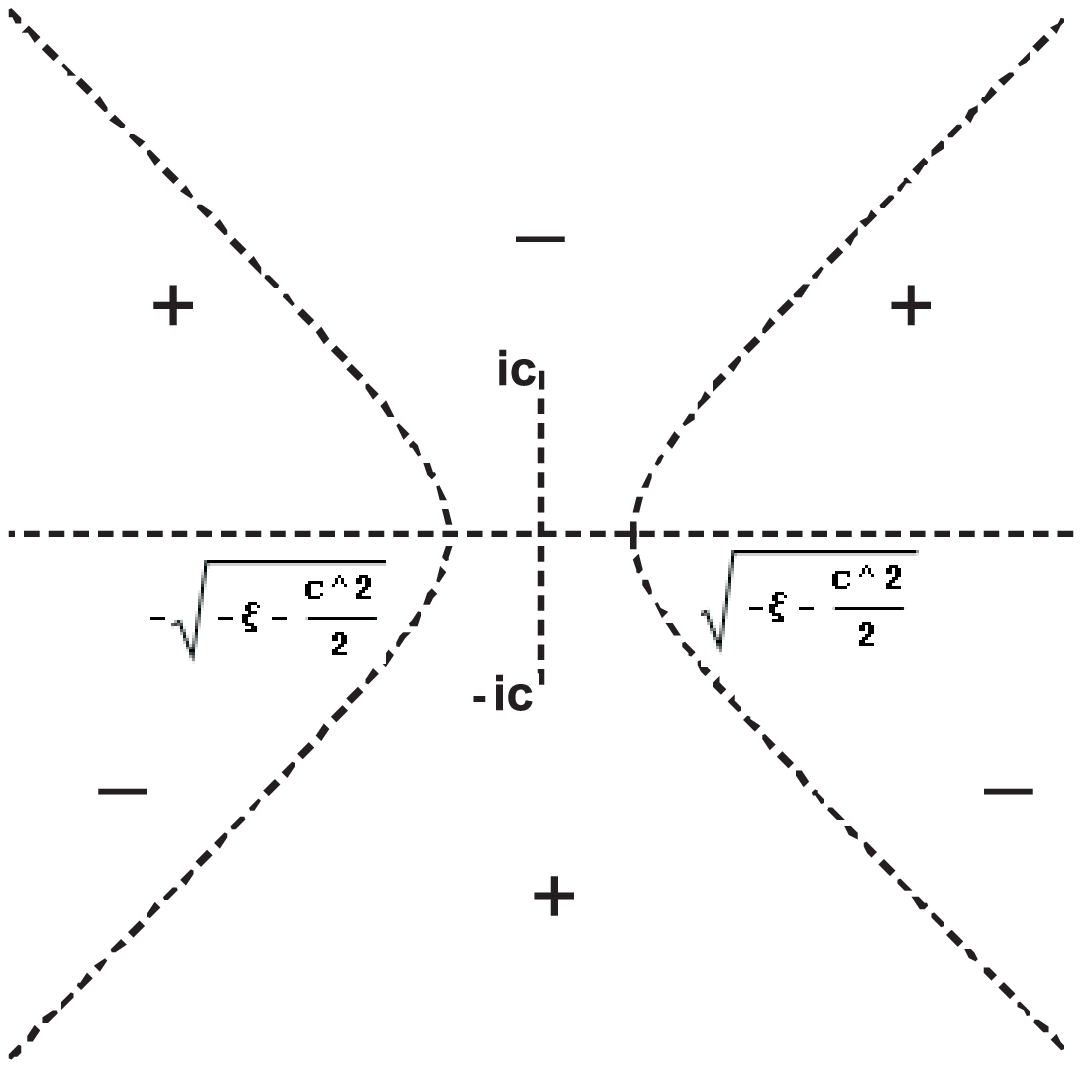}
\end{center}
\caption{The signature table of the function $\Im g_c(k,\xi)$
($\xi<\xi_-$)} \label{im-g3}
\end{figure}

\begin{figure}[ht]
\begin{center}
\epsfig{width=50mm,figure=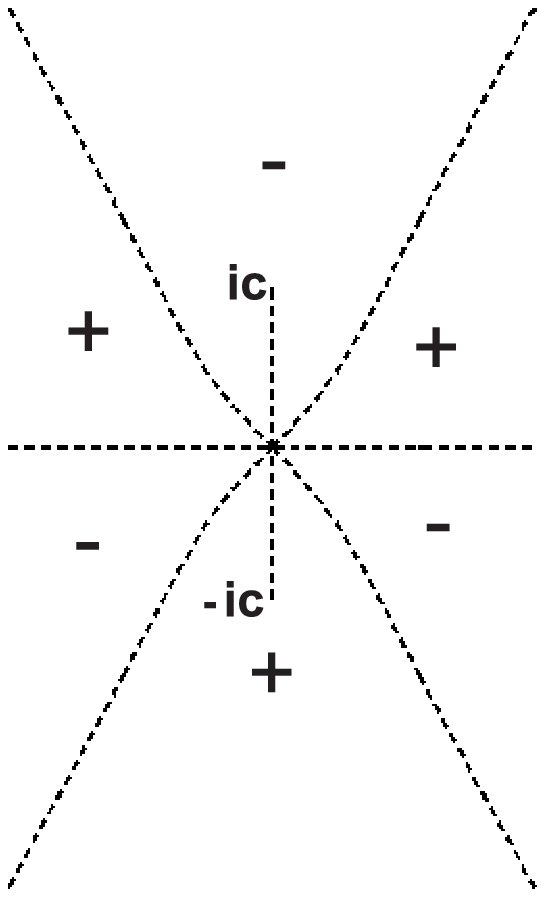}
\end{center}
\caption{The signature table of the function $\Im  g_c(k,\xi)$
($\xi=\xi_-$)} \label{im-g4}
\end{figure}

In what follows very important role plays a signature table of
the function $\Im g_c(k,\xi)$ for different values of $\xi$.
Borderlines between different domains are described by equations:
\[k_2=0; \qquad\qquad k_1=0\qquad \rm{and} \qquad |k_2|\le c; \]
\[
3\left(k_1^2-k_2^2+\xi+\frac{c^2}{2}\right)\left(k_1^2-k_2^2+3\xi-
\frac{c^2}{2}\right)= 4k_1^2k_2^2,
\]
which are equivalent to $\Im g_c(k,\xi)=0$. The signature table
of the function $\Im  g_c(k,\xi)$ can be obtain by using for
example "MAPLE" and it is qualitatively depicted on the figure
\ref{im-g3} for $-\infty<\xi<\xi_-$ and on the figure \ref{im-g4}
for $\xi=\xi_-$.

Thus  $g_c(k,\xi) $ as the function on $k$ has the following
properties:
\begin{itemize}
  \item $g_c(k,\xi)$ is analytic in $\mathbb{C}\backslash[-ic,ic]$
  \item $g_c(k,\xi)=\theta(k,\xi)+ \ord\left( k^{-1}\right),\quad k\rightarrow\infty$
  \item $g_c^+(k,\xi)+g_c^-(k,\xi)=0,\quad k\in(-ic,ic)$
  \item $g_c(k,\xi)=\ord(\sqrt{k\mp ic})$,\quad   $k\to\pm ic$.
\end{itemize}

The Riemann-Hilbert problem for the matrix $M(\xi,t,k)$ with the
jump contour $\Sigma=\Db{R}\cup [ic, -ic] $ have to be considered
now with new (\ref{gpw}) phase function $ g_c(k,\xi)$. Let us
define new matrix function
$$
M^{(1)}(\xi,t,k)=M(\xi,t,k)G^{(1)}(\xi,t,k),
$$
where
$G^{(1)}(\xi,t,k)=e^{it(g_c(k,\xi)-\theta(k,\xi))\sigma_3}.$ The
function $M^{(1)}$ solve the following R-H problem:
\[
M^{(1)}_-(\xi,t,k)=M^{(1)}_+(\xi,t,k)J^{(1)}(\xi,t,k), \qquad
M^{(1)}(\xi,t,k)\rightarrow I,\quad k\rightarrow\infty
\]
where
\[
J^{(1)}(\xi,t,k)=\begin{cases}
\left(\begin{array}{ccc}1&-\overline{r(k)}e^{-2itg_c(k,\xi)}\\\\
-r(k)e^{2itg_c(k,\xi)}&1+|r(k)|^2\end{array}\right),\quad k\in
\mathbb{R}\\\\
\left(\begin{array}{ccc}1&0\\\\f(k)e^{2itg_c(k,\xi)}&1\end{array}
\right),\quad k\in(0,ic)\\\\
\left(\begin{array}{ccc}1&f(k)e^{-2itg_c(k,\xi)}\\\\0&1\end{array}
\right),\quad k\in(-ic,0)
\end{cases}
\]
Further we would like to transfer the jump contour from the real
axis. To do so we use the following factorizations of the jump
matrix on the real axis:
\begin{align} \label{flu1}
J^{(1)}(\xi,t,k)&=
\left(\begin{array}{ccc}1&0\\\\
-r(k)e^{2itg_c(k,\xi)}&1\end{array}\right)
\left(\begin{array}{ccc}1&-\overline{r(k)}e^{-2itg_c(k,\xi)}\\\\
0&1\end{array}\right)\\&= \left(\begin{array}{ccc}1&
\displaystyle\frac{-\overline{r(\overline{k})}e^{-2itg_c(k,\xi)}}
{ 1+r(k)\overline{r(\overline{k})} }
\\\\0&1\end{array}\right)
\left(\begin{array}{ccc}\displaystyle\frac{1}{1+|r(k)|^2}&0\\\\0&1+|r(k)|^2
\end{array}\right)
\left(\begin{array}{ccc}1&0\\\\\displaystyle\frac{-r(k)e^{2itg_c(k,\xi)}}
{1+r(k)\overline{r(\overline{k})}}&1
\end{array}\right)\label{flu2}
\end{align}
 It is easy to see, the first(second) factor in the first line (\ref{flu1}) is decrease as
$t\rightarrow\infty$ in the domains where $\mathrm{Im}\
g_c(k,\xi)>0$ $ (\mathrm{Im}\ g_c(k,\xi)<0)$. In the second line
(\ref{flu2}) the first(third) factor is decrease as
$t\rightarrow\infty$ in the domains where $\mathrm{Im}\
g_c(k,\xi)<0$ $ (\mathrm{Im}\ g_c(k,\xi)>0)$. To remove the
diagonal terms in the second factorization we use a diagonal
transformation:
\[M^{(2)}(\xi,t,k)=M^{(1)}(\xi,t,k)\delta^{-\sigma_3}(k,\xi),\qquad
\delta^{-\sigma_3}(k,\xi)=\begin{pmatrix}\delta^{-1}(k,\xi)&0\\
0&\delta(k,\xi)\end{pmatrix},
\]
where some analytic in
$\mathbb{C}\backslash[-\lambda(\xi),\lambda(\xi)]$ function
$\delta(k,\xi)$ must be defined. Then the jump matrix
$J^{(2)}(\xi,t,k)$ for $k\in [-\lambda(\xi), \lambda(\xi)]$ takes
the form
\[
J^{(2)}(\xi,t,k)=
\begin{pmatrix}1&\displaystyle\frac{-\overline{r }
\delta_+^2 e^{-2itg_c(k,\xi)}}{1+|r|^2 }\\\\0&1
\end{pmatrix}
\begin{pmatrix}\displaystyle\frac{\delta_+ }{\delta_-}
(1+|r |^2)^{-1}&0\\
0&\displaystyle\frac{\delta_-}{\delta_+ }
\left(1+|r|^2\right)\end{pmatrix}
\begin{pmatrix} 1&0\\\\\displaystyle\frac{-r e^{2itg_c(k,\xi)}}
{ (1+|r|^2 ) \delta_-^2 }&1\end{pmatrix}.
\]
If we chose the function $\delta(k,\xi)$ in the form:
\begin{equation}\label{delta}\delta(k,\xi)=\left(\displaystyle\frac{k-
\lambda(\xi)}{k+
\lambda(\xi)}
\right)^{-i\nu}\chi(k,\xi),\end{equation}
where
\begin{equation}\label{chi}\chi(k,\xi)=\exp\left(\displaystyle
\frac{1}{2\pi
i}\displaystyle\int\limits_{-\lambda(\xi)}^{\lambda(\xi)}\displaystyle\frac
{\ln\left(\displaystyle\frac{1+|r(s)|^2}
{1+|r(d)|^2}\right)ds}{s-k}\right)\end{equation} and
\begin{equation}\label{nu}\nu=\displaystyle\frac{1}{2\pi}\ln(1+
|r(\lambda(\xi))|^2)
\end{equation}then
$\delta_+=\delta_-(1+|r|^2)$ and, therefore,  the middle matrix
factor become trivial. Thus the jump matrix $J^{( 2)}(\xi,t,k)$
has a lower/upper factorization for $k\in [-\lambda (\xi),
\lambda (\xi)]$ and a upper/lower factorization for $k\notin
[-\lambda (\xi), \lambda (\xi)]$:
\[
J^{(2)}(\xi,t,k)=\begin{cases}
\begin{pmatrix}
1& a(k)b(k)\delta^{2}_+(k,\xi)\ee^{-2\ii t g_c(k,\xi)}\\0&1
\end{pmatrix}
\begin{pmatrix}
1&0\\-a(k)b(k)\delta^{-2}_-(k,\xi)\ee^{2\ii t g_c(k,\xi)}&1
\end{pmatrix},\\ \hfill k\in [-\lambda(\xi), \lambda(\xi)]\\  \\
\begin{pmatrix}
1&0\\-r(k)\delta^{-2}(k,\xi)\ee^{2\ii t g_c(k,\xi)}&1
\end{pmatrix}
\begin{pmatrix}
1&r(k)\delta^{2}(k)\ee^{-2\ii t g_c(k,\xi)}\\0&1
\end{pmatrix},\qquad k\notin [-\lambda(\xi), \lambda(\xi)],
 \end{cases}
\]
where we use the identity:
\[
\frac{r(k)}{1+|r(k)|^2}=a(k)b(k).
\]
For $k\in[-ic,ic]$ the jump matrix $J^{(2)}(\xi,t,k)$ takes the
form:
\begin{align*}
\begin{pmatrix}
\ee^{- \ii t(g^+_c(k,\xi)-g^-_c(k,\xi))}&0\\f(k)\ee^{\ii
t(g^+_c(k,\xi)+g^-_c(k,\xi))}& \ee^{\ii
t(g^+_c(k,\xi)-g^-_c(k,\xi))}
\end{pmatrix},
& \qquad k\in[0,ic], \\
\begin{pmatrix}
\ee^{- \ii t(g^+_c(k,\xi)-g^-_c(k,\xi))}& f(k)\ee^{-\ii
t(g^+_c(k,\xi)+g^-_c(k,\xi))}\\0& \ee^{\ii
t(g^+_c(k,\xi)-g^-_c(k,\xi))}
\end{pmatrix},
&\qquad k\in [-ic, 0].
\end{align*}
To remove unbounded in $t$ exponentials we have to put
\begin{equation}\label{g=0}
g_c^-(k,\xi)+g_c^+(k,\xi)=0, \qquad k\in [-ic, ic] .
\end{equation}
Chosen above function $g_c(k,\xi)$ satisfies this condition and
the jump matrix takes the form
\begin{align*}
J^{(2)}(\xi,t,k)&=\begin{pmatrix}
\ee^{-2 \ii t g_c^+(k,\xi)}&0\\
f(k)\delta^{-2}(k)&\ee^{2 \ii t  g_c^+(k,\xi)}
\end{pmatrix},\qquad k\in[ic,0]\\ \\
&=\begin{pmatrix}
\ee^{-2 \ii t  g_c^+(k,\xi)}& f( k)\delta^{2}(k)\\
0&\ee^{2 \ii t  g_c^+(k,\xi)}
\end{pmatrix},\qquad k\in[0,ic]
\end{align*}
on the contour $[ic, -ic]$. The jump contour $\Sigma^{(2)}$ for
$M^{( 2)}(\xi,t,k)$-problem is the  initial one, i.e.
$\Sigma^{(2)}=\Sigma $.

\begin{figure}[ht]
\begin{center}
\epsfig{width=100mm,figure=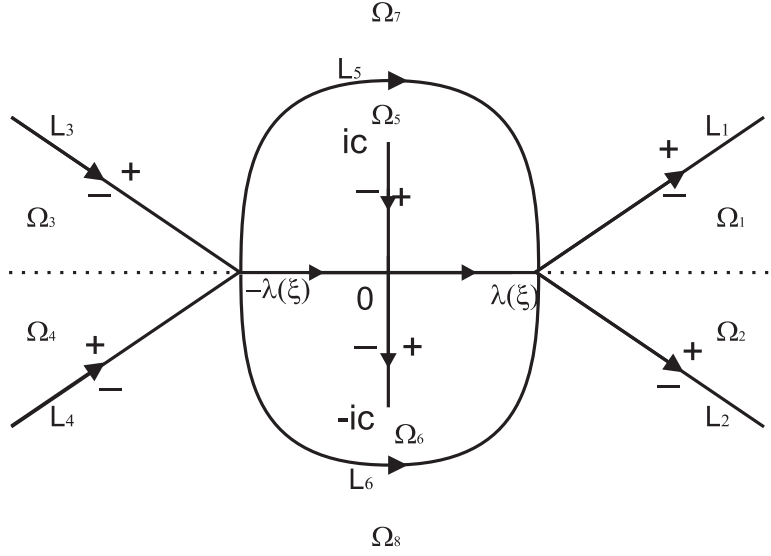}
\end{center}
\caption{The contour $\Sigma^{(3)}$ for the
$M^{(3)}(\xi,t,k)$-problem} \label{Contoursteady1}
\end{figure}

Let us define a decomposition of the complex $k$-plane into eight
domains $ \Omega_1, ...., \Omega_8$ separated by their common
boundary $\Sigma^{(3)}$ as it is shown on the figure \ref{Contoursteady1}.
The contours $L_1$, $L_6$ ($L_3$, $L_4$) range from the point
$\lambda(\xi)$ ($-\lambda(\xi$)) to infinity along the rays $\arg
k=\pm\pi/4$ ($\arg k=\pi\mp\pi/4$);  the contours $L_2$ and
$L_5$, passing through the points $\pm\lambda(\xi)$, form a
closed oval containing segment $[-ic,ic]$. Then the next
transformation is:
\[
M^{(3)}(\xi,t,k) = M^{(2)}(\xi,t,k) G^{(2)}(\xi, t,k),
\]
where
\begin{align}
\label{G} & G^{(2)}(\xi, t,k)=
\begin{cases}
\begin{pmatrix} 1&0\\-r(k)\delta^{-2}(k,\xi)\ee^{2 \ii tg_c(k,\xi)}&
1\end{pmatrix},
& k\in\Omega_1\cup\Omega_3, \\ \\
\begin{pmatrix} 1&0\\0&1
\end{pmatrix},
& k\in\Omega_2\cup\Omega_5, \\ \\
\begin{pmatrix} 1&-r(k)\delta^{2}(k,\xi)\ee^{-2 \ii tg_c(k,\xi)} \\
0&1\end{pmatrix},& k\in\Omega_4\cup\Omega_6 ,
\end{cases}
\end{align}
\begin{align}\label{G1}
& G^{(2)}(\xi, t,k)=
\begin{cases}
\begin{pmatrix} 1&a(k)b(k)\delta^{2}(k,\xi)\ee^{-2 \ii t g_c(k,\xi)}\\0&1
\end{pmatrix},
& k\in\Omega_7, \\ \\
\begin{pmatrix} 1&0\\a(k)b(k)\delta^{-2}(k,\xi)
\ee^{2 \ii t g_c(k,\xi)}&1\end{pmatrix},& k\in\Omega_8.
\end{cases}
\end{align}

The $G^{(2)}$-transformation  gives to the following RH problem:
\[
M^{(3)}_-(\xi,t,k)=M^{(3)}_+(\xi,t,k)J^{(3)}(\xi,t,k), \qquad
M^{(3)}_-(\xi,t,k)=I+\ord(k^{-1}), \quad k\to\infty
\]
\noindent on the contour $\Sigma^{(3)}$ depicted on the Figure
\ref{Contoursteady1} with jump matrices $J^{(3)}(\xi,t,k)$ which are equal to
the identity matrix on the real axis, they coincide with matrices
$G^{(2)}(k)$ from (\ref{G})-(\ref{G1}) written for the contours
$k\in L_j$ ($j=1,2,...,6$). It is easy to see that
$J^{(3)}(\xi,t,k)=I+O(\ee^{-\epsilon t})$ as $t\to\infty$ and
$k\in L_j$ with exception of some neighborhoods of the stationary
points $\pm\lambda(\xi)$. Therefore the main  contribution  to
the asymptotics comes from  the jump matrix on the segment $[ic,
-ic]$, where it takes the form:

$J^{(3)}(\xi,t,k)={G^{(2)}_+}^{-1}(k)J^{(2)}(\xi,t,k){G^{(2)}_-}(k)=$
\begin{align*}
=&\begin{cases}\begin{pmatrix}
1&-a_+(k)b_+(k)\delta^2(k,\xi)\ee^{-2 \ii tg_c^+(k,\xi)}\\
0&1
\end{pmatrix}
J^{(2)}(\xi,t,k)
\begin{pmatrix}
1&a_-(k)b_-(k)\delta^{2}(k,\xi)\ee^{2 \ii tg_c^+(k,\xi)}\\0&1
\end{pmatrix},\\
\hfill k\in[0,ic], \\ \\
\begin{pmatrix}
1&0\\-a_+(k)b_+(k)\delta^{-2}(k,\xi)\ee^{2 \ii tg_c^+(k,\xi)}&1
\end{pmatrix}
J^{(3)}(\xi,t,k)
\begin{pmatrix}
1&0\\a_-(k)b_-(k)\delta^{-2}(k,\xi)\ee^{-2 \ii tg_c^+(k,\xi)}&1
\end{pmatrix},\\
\hfill k\in[0,-ic].
\end{cases}
\end{align*}
Using equalities $1-f(k)a_+(k)b_+(k)=0$, $1-f(k)a_+(k)b_+(k)=0$
and $a_+(k)b_+(k)=-a_-(k)b_-(k)$, which follow from the
definition of the function $f(k)$, we obtain
\[
J^{(3)}(\xi,t,k)=\begin{cases}\begin{pmatrix}
0&- f^{-1}(k)\delta^2(k,\xi)\\
f(k)\delta^{-2}(k,\xi)&0
\end{pmatrix}, \qquad k\in[0,ic]\\\\
\begin{pmatrix}
0& f(k)\delta^{2}(k,\xi)\\
- f^{-1}(k)\delta^{-2}(k,\xi)&0
\end{pmatrix},\qquad k\in[0,-ic].\end{cases}
\]

 Further we would like to obtain the RH problem with
constant in $k$ jump matrix. To do so let us use the
factorization
\begin{align*}
J^{(3)}(\xi,t,k) =
\begin{pmatrix}
F^{-1}_+(k,\xi)&0\\0&F_+(k,\xi)
\end{pmatrix}
\begin{pmatrix}
0&\ii \\ \ii &0
\end{pmatrix}
\begin{pmatrix}
F_-(k,\xi)&0\\0&F^{-1}_-(k,\xi)
\end{pmatrix}
\end{align*}
which takes place if $F_-(k,\xi)F_+(k,\xi)=-\ii
f(k)\delta^{-2}(k,\xi)$ for $ k\in [0,ic]$ and
$F_-(k,\xi)F_+(k,\xi)=\ii f^{-1}(k,\xi)\delta^{-2}(k,\xi)$ for $
k\in[0,-ic]$.

Thus  we come to the scalar Riemann-Hilbert problem: find  a
scalar function $ F(k,\xi)$ such that
\begin{itemize}
\item $ F(k,\xi)$ is analytic outside the contour $ [ic,-ic]$
which is oriented from $ic$ to $-ic$.; \item $ F(k,\xi)$ does not
vanish; \item $ F(k,\xi)$ satisfies the jump relation:
\[
  F_-(k,\xi)  F_+(k,\xi)=h(k)\delta^{-2}(k,\xi), \qquad
k\in [ic,-ic],
\]
where
\[
h(k)=\begin{cases}-\ii f(k)=a^{-1}_-(k)a^{-1}_+(k) \quad k\in [0,ic], \\
\ii  f^{-1}(k)=a_-(k)a_+(k), \quad k\in[0,-ic].
\end{cases}
\]
\item $ F(k,\xi)$ is bounded at infinity.
\end{itemize}
To solve this RH problem let us put
\begin{equation}\label{F}F(k,\xi)=\begin{cases}\displaystyle\frac{1}{a(k)}
F_{aux}(k,\xi),
\quad k\in
\mathbb{C}_+\backslash[0,ic]\\\\a(k)F_{aux}(k,\xi),\quad k\in
\mathbb{C}_-\backslash[-ic,0]\end{cases},\end{equation} and use
the function $X(k)=\sqrt{k^2+c^2 }$. Since
\[
\left[\frac{\log {F_{aux}(k,\xi)}}{X(k)}\right]_+ -
\left[\frac{\log{F_{aux}(k,\xi)}}{X(k)}\right]_-
=\frac{\log{\delta^{-2}(k)}}{X_+(k)}, \qquad k\in   [ic,-ic],
\]
and
\[
\left[\frac{\log {F_{aux}(k,\xi)}}{X(k)}\right]_+ -
\left[\frac{\log{F_{aux}(k,\xi)}}{X(k)}\right]_-
=\frac{\log{a^{2}(k)}}{X(k)}, \qquad k\in\Db{R},
\]
we have
\begin{equation}\label{Faux}
F_{aux}(k,\xi)=\exp\left\{\frac{X(k)}{2\pi\ii}\left[
\int\limits_{\Db{R}} \frac{\log a^{2}(s)}{s-k}\frac{ds}{X(s)}-
\int\limits_{-ic}^{ic}
\frac{\log\delta^{-2}(s,\xi)}{s-k}\frac{ds}{X_+(s)}
\right]\right\}.
\end{equation}
Let us note, that $F(\infty,\xi)=1$.

\noindent Indeed, as $a(\infty)=1$ then $F(k,\infty)=F_{aux}(k,\infty)$
\begin{equation}\label{Fauxinfty}
F_{aux}(\infty,\xi)=\exp\left\{\frac{-1}{2\pi\ii}\left[
\int\limits_{\Db{R}} \log a^{2}(s)\frac{ds}{X(s)}-
\int\limits_{-ic}^{ic}
\log\delta^{-2}(s,\xi)\frac{ds}{X_+(s)}
\right]\right\}.
\end{equation}
\noindent As
\begin{itemize}
    \item for $s\in \mathbb{R}\quad a(s)>0\quad\&\quad a(-s)=a(s)$
    \item for $s\in(-ic,ic) \quad\delta(s,\xi)>0\quad\&\quad\delta(-s,\xi)=1/\delta(s,\xi)$
    \item for $s\in(-ic,ic)\quad X_+(-s)=X_+(s)$
\end{itemize}
\noindent then
$\displaystyle\int\limits_{\Db{R}} \log a^{2}(s)\frac{ds}{X(s)}=0$
and
$\displaystyle\int\limits_{-ic}^{ic}
\log\delta^{-2}(s,\xi)\frac{ds}{X_+(s)}=0$

\noindent Finally, $F(\infty,\xi)=1.$

\noindent Now, since
\[
J^{(3)}(\xi,t,k)=  F^{-\sigma_3}_+(k,\xi)J^{mod}
F^{\sigma_3}_-(k,\xi), \qquad k\in [ic, -ic],
\]
where
\begin{align*}
J^{mod}=&\begin{pmatrix} 0&\ii \\ \ii &0
\end{pmatrix}
\end{align*}
the next  step  is as follows:
\[
M^{(4)}(\xi,t,k)=  M^{(3)}(\xi,t,k) F^{-\sigma_3}(k,\xi), \qquad
\]
Then we have
\[
M_-^{(4)}(\xi,t,k) =M_+^{(4)}(\xi,t,k)J^{(4)}(\xi,t,k), \qquad
k\in\Sigma_4 ,
\]
where
\begin{align*}
J^{(4)}(\xi,t,k)=\begin{cases}
I &k\in \Db{R},\\
J^{mod}       &k\in (ic,-ic), \\
I+\ord(e^{-\varepsilon t})&k\in L_j,\quad j=1,2,...,6.
\end{cases}
\end{align*}

 The analysis of the parametrix solutions near
the end points $ic$, $ -ic$ and the stationary points
$\pm\lambda(\xi)$ are very similar to the analysis done in
\cite{DIZ} and \cite{DIZ93}, respectively. In the first case,
since the local representation of $g_c(k,\xi)$ at the points $ic$
and $-ic$ is characterized by a square root type behavior:
\[
g_c(k,\xi)\sim g_0(ic,\xi)\sqrt{k-ic},\quad k\to ic;\qquad
g_c(k,\xi)\sim\bar g_0(-ic,\xi)\sqrt{k+ic}, \quad k\to-ic,
\]
the relevant model Riemann-Hilbert problems are solvable in terms
of the Bessel functions while in the second case of the real
stationary points $\pm\lambda(\xi)$ they are solvable in terms of
the parabolic cylinder functions. The contribution to the
asymptotics has the order $\ord(t^{-3/2})$ ?,
$\ord(e^{-\epsilon\sqrt t})$ ? in the first case and
$\ord(t^{-1/2})$ in the second case. Therefore we have:
\begin{equation}\label{m4final}
M^{(4)}(\xi,t,k)=\left( I +
O\left(\frac{1}{t^{1/2}}\right)\right)M^{(mod)}(k),
\end{equation}
where $M^{(mod)}(k)$ solves the zero-gap model problem (cf.
\cite{BIK}):
\[
M^{(mod)}_-(k)=M^{(mod)}_+(k)J^{(mod)} \qquad k\in(ic,
-ic),\qquad M^{(mod)}(k)=I+\ord(k^{-1}), \quad k\to\infty.
\]
with constant jump matrix:
\[
J^{mod}=\begin{pmatrix} 0&\ii \\ \ii &0
\end{pmatrix}.
\]

To solve the model problem let us  use the function
\[
\varkappa(k)=\sqrt[4]{\displaystyle\frac{k-ic}{k+ic}}
\]
introduced in the first section. Since
$\varkappa_-(k)=\ii\varkappa_+(k)$ on the cut $(ic, -ic)$ the
explicit solution of the model problem takes the form:
\[
M^{(mod)}(k)=\frac{1}{2}
\begin{pmatrix}
\varkappa(k)+\DS\frac{1}{\varkappa(k)}&
\varkappa(k)-\DS\frac{1}{\varkappa(k)}\\
\varkappa(k)-\DS\frac{1}{\varkappa(k)}&
\varkappa(k)+\DS\frac{1}{\varkappa(k)}
\end{pmatrix}
\]

Finally we have the following chain of transformations of the RH
problem:
\begin{align*}
&M(\xi,t,k)= M^{(1)}(\xi,t,k)\ee^{\ii t[\theta(k)-g_c(k,\xi)]
 \sigma_3},\\
&M^{(1)}(\xi,t,k)=M^{(2)}(\xi,t,k)\delta^{\sigma_3}(k,\xi), \\
&M^{(2)}(\xi,t,k)=M^{(3)}(\xi,t,k)[G^{(2)}(\xi,t,k)]^{-1}\\
&M^{(3)}(x,t,k)=  M^{(4)}(\xi,t,k) F^{\sigma_3}(k,\xi),\\
&M^{(4)}(x,t,k)=M^{(mod)}(k)(I+O(t^{-1/2})).
\end{align*}
Let us emphasize that any matrix $M^{(j)}(\xi,t,k)$ ($j=1,2,...
$) defines the same functions $q(x,t)$ since all bordering
matrices are diagonal at the point $k=\infty$. By the Theorem 2.1
$q(x,t)=2\ii\lim\limits_{k\to\infty}[kM(x/12t,t,k)]_{12} $. If we
denote
$\lim\limits_{k\to\infty}[kM^{(j)}(x/12t,t,k)]_{12}=m^{(j)}_{12}(x,t)
$ then, take into account the chain of our transformations and
using the equalities $m^{mod}_{12}(x,t)= c/2\ii$,
$F(\infty,\xi)=1$, we have:
\begin{align*}
q(x,t)&=2\ii m_{12}(x,t)=2\ii m^{(1)}_{12}(x,t)=2\ii
  m^{(2)}_{12}(x,t)\\
&=2\ii  m^{(3)}_{12}(x,t)+\ord(t^{-1/2})\\
&=2\ii  m^{(4)}_{12}(x,t)F^{-2}(\infty,\xi)+\ord(t^{-1/2})\\
&=2\ii m^{mod}_{12}(x,t)
F^{-2}(\infty,\xi)+\ord(t^{-1/2})\\
&= c+\ord(t^{-1/2}).
\end{align*}

\begin{teor} The solution of the IBV problem (\ref{mkdv})-(\ref{iv})
for $t\to\infty$ in the region $-\infty<x<-6c^2t$ takes the form:
\begin{eqnarray*}
  q(x,t) &=& c + O(t^{-1/2}),  \\
\end{eqnarray*}
\end{teor}


\subsection{Elliptic-wave region $-6c^2t<x<4c^2t$}

\subsubsection{The construction of the $g$-function}

We need a function $g(k,\xi)$ with the following properties:
\begin{enumerate}
  \item g is analytic against k in $\mathbb{C}\backslash[-ic,ic]$
  \item $\exists\lim\limits_{k\rightarrow\infty}(g(k,\xi)-\theta(k,\xi))\in
      \mathbb{C}$
  \item a set of points where $\mathrm{Im}\ g(k,\xi)=0$ divides the complex plane into four
  connected open sets and contains the set $\mathbb{R}\bigcup[ic,id]\bigcup[-ic,-id]$,
  where $d\in(0,c)$ is some number.
\end{enumerate}
We will find such a function in the  form:
\[
g(k,\xi)=\int\limits_{ic}^k\frac{12(s^2+\mu^2)(s^2+d^2)ds}
{\mathrm{w}(s)},\qquad \mathrm{w}(s)=\sqrt{(s^2+c^2)(s^2+d^2)},
\]
where the positive on the real axis $\mathbb{R}$ function
$\mathrm{w}(s)$ is analytic in $\mathbb{C}\backslash[-ic,ic]$ and
numbers $\mu$ and $d$ have to be determined as functions on
$\xi$. The integration contour is taken to have no intersection
with the segment $(-ic,ic)$. It is easy to see that
$g(k,\xi)\in\mathbb{R}$ if $k\in[c,d]_+$ or $k\in[c,d]_-$. To
satisfy the requirement $g(k,\xi)\in\mathbb{R}$ if
$k\in[-c,-d]_+$ or $k\in[-c,-d]_-$ we have to choose the numbers
$\mu$ and $d$ in such a way that $\int\limits_{-ic}^{ic}dg(k)=0$.
Besides this requirement makes $g(k,\xi)$ being a single-valued
function on $\mathbb{C}\setminus[-ic,ic]$ against $k$. The
mentioned requirement can be written as follows:
\begin{equation}\label{F=0}
\int\limits_0^1\left(\mu^2-\lambda^2d^2\right)\sqrt{
 {\frac{1-\lambda^2}{c^2-\displaystyle{\lambda^2d^2}}}}d\lambda=0.
\end{equation}

Let us  define the function
\begin{equation}\label{Fmud}
F(\mu,d)=\int\limits_0^1\left(\mu^2-\lambda^2d^2\right)\sqrt{
 {\frac{1-\lambda^2}{c^2-\displaystyle{\lambda^2d^2}}}}d\lambda.
\end{equation}
It is easy to see that there  exists a function $\mu=\mu(d)$ such
that $F(\mu(d),d)\equiv0$ and $0<\mu(d)<d$. Moreover, $\mu(d)$ is
strictly increasing when $d\in[0,c]$. Indeed, one can check that
$F(\mu,d)$ is strictly increasing against $\mu$ and is strictly
decreasing against $d$ when $0<\mu<d<c$. Now if $0<d_1<d_2<c$
then $F(\mu(d_1),d_1)=0=F(\mu(d_2),d_2)<F(\mu(d_2),d_1)$, that is
$F(\mu(d_1),d_1)<F(\mu(d_2),d_1)$ and, hence,
$\mu(d_1)<\mu(d_2)$.  Furthermore $\mu(d)$ is continuous
function. It follows from \ref{F=0}:
\begin{equation}\label{mud}
\mu^2(d)= \int\limits_0^1 \lambda^2d^2\sqrt
 {\frac{1-\lambda^2}{c^2-\lambda^2d^2}}d\lambda\Bigg/
\int\limits_0^1  \sqrt
 {\frac{1-\lambda^2}{c^2-\lambda^2d^2}}d\lambda
\end{equation}
Now we want to satisfy the requirement (2)
$\exists\lim\limits_{k\rightarrow\infty}(g(k,\xi)-\theta(k,\xi))\in
\mathbb{C}$. For this enough that $dg(k,\xi)-d\theta(k,\xi)= \ord
(k^{-2} ) \quad {\rm as} \quad k\rightarrow\infty$. Since
$d\theta(k,\xi)=12(k^2+\xi)dk$ and
$dg(k,\xi)=[12(k^2+\mu^2-\displaystyle\frac{c^2}{2}+
\displaystyle\frac{d^2}{2})+\ord (k^{-2} )]dk  \quad {\rm as}
\quad k\rightarrow\infty$,  we require that
\begin{equation}\label{cximud}
\displaystyle\frac{c^2}{2}+\xi=\mu^2+
\displaystyle\frac{d^2}{2}.
\end{equation}
 Equation (\ref{mud}) yields that $\mu(0)=0$ and
$\mu(c)=\displaystyle\frac{c}{\sqrt{3}}.$  Hence
$\mu^2(d)+\displaystyle\frac{d^2}{2}$ is vary over the segment
$[0,\displaystyle\frac{5c^2}{6}]$ when $d$ is vary over the
segment $[0,c]$. So for any
$\xi\in\left(-\displaystyle\frac{c^2}{2},\displaystyle\frac{c^2}{3}
\right)$ there exists a single $d=d(\xi)\in(0,c)$ such that
\begin{equation}\label{cximud2}\displaystyle\frac{c^2}{2}+\xi=
\mu^2(d(\xi))+ \displaystyle\frac{d(\xi)^2}{2}.\end{equation} It
is easy to see from (\ref{cximud2}) that $d(\xi)$ is continuous
as the function on
$\xi\in\left(-\displaystyle\frac{c^2}{2},\displaystyle\frac{c^2}{3}
\right)$. Thus the function $g(k,\xi )$ is completely defined.

\begin{figure}[ht]
\begin{center}
\epsfig{width=80mm,figure=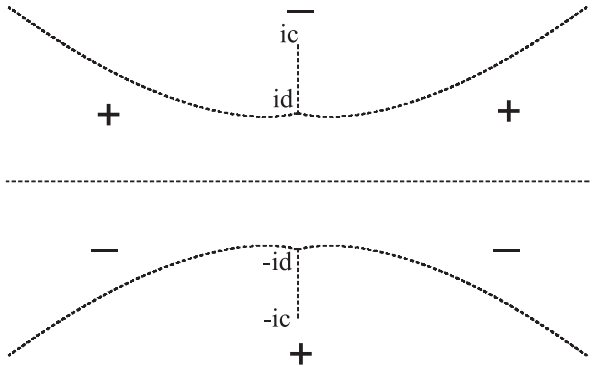}
\end{center}
\caption{The signature table for Im g} \label{Im g=0}
\end{figure}

The signature table of the imaginary part of the function
$g(k,\xi)$ is depicted on figure \ref{Im g=0}.

The function $g(k,\xi)$ has the following additional properties:
\begin{enumerate}
\setcounter{enumi}{3}
  \item[2.a] $\lim\limits_{k\rightarrow\infty}\left(g(k,\xi)-
  \theta(k,\xi)\right)=0$\\
  It is follows from the facts that this limit exists and $$g(k,\xi)-\theta(k,\xi)\in \ii\mathbb{R}\quad ,\quad k\in(\\ic,i\infty)$$and
  $$g(k,\xi)-\theta(k,\xi)\in \mathbb{R}\quad ,\quad k\in\mathbb{R}$$
  \item $g_-(k,\xi)+g_+(k,\xi)=0,\quad
  k\in(ic,id)\bigcup(-ic,-id)$;
  \item $g_-(k,\xi)-g_+(k,\xi)=B_g(\xi)>0,\quad k\in(-id,id)$,
  where
  $$
B_g(\xi)=2\int\limits_{id}^{ic} dg(k,\xi)= 24 \int\limits_d^c
\left[\left(s^2-\mu^2(\xi)\right)\sqrt{\displaystyle\frac
  {s^2-d^2(\xi)}{c^2-s^2}}\right]ds=
  $$$$\hfill=24\displaystyle\int\limits_{\frac{d}{c}}
  ^1\frac{\left(t^2-\frac{\mu^2(\xi)}{c^2}\right)\sqrt{t^2-\frac{d^2(\xi)}
  {c^2}}}{\sqrt{1-t^2}}dt.
  $$
\end{enumerate}

\subsubsection{Transition from initial R-H problem to the model one.}

\begin{figure}[ht]
\begin{center}
\epsfig{width=100mm,figure=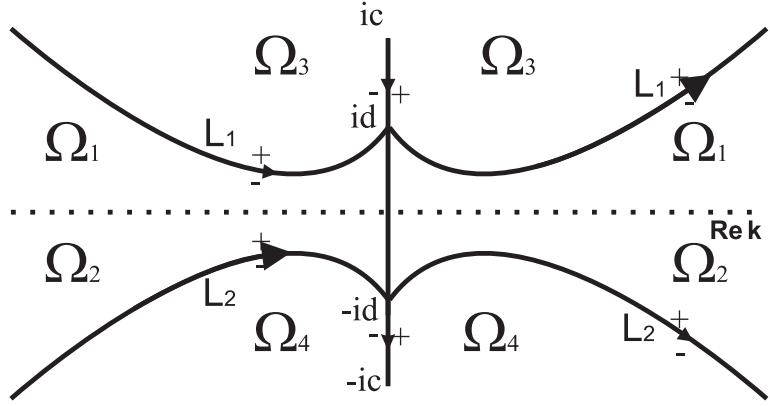}
\end{center}
\caption{The contour $\Sigma_2$ for the R-H problem $M^{(2)}(\xi,t,k)$}
\label{Contourelliptic1}
\end{figure}

Now we introduce the chain of the transformations of the R-H
problem.
\\\textbf{Step 1}\\
First, we change the phase function $\theta(k,\xi)$ with
$g(k,\xi)$:
\[M^{(1)}(\xi,t,k)=M(\xi,t,k)e^{it(g(k,\xi)-
\theta(k,\xi))\sigma_3}.\] Then \[M^{(1)}(\xi,t,k)\rightarrow
I,\quad k\rightarrow\infty\]and\[M^{(1)}_-(\xi,t,k)=
M^{(1)}_+(\xi,t,k)J^{(1)}(\xi,t,k),\] where
\begin{align*}
J^{(1)}(\xi,t,k)&=
\left(\begin{array}{ccc}1&-\overline{r(k)}e^{-2itg(k,\xi)}\\\\
-r(k)e^{2itg(k,\xi)}&1+|r(k)|^2\end{array}\right),
&k\in\mathbb{R}
\end{align*}
\begin{align*}
&=\left(\begin{array}{ccc}e^{itB_g(\xi)}&0\\\\f(k)
e^{it(2g_+(k,\xi)+B_g(\xi))}&e^{-itB_g(\xi)}
\end{array}\right), & k\in(0,id)\end{align*}
\begin{align*}
&=\left(\begin{array}{ccc}e^{itB_g(\xi)}&f(k)
e^{-it\left(2g_+(k,\xi)+B_g(\xi)\right)}\\\\0&e^{-itB_g(\xi)}
\end{array}\right),& k\in(0,-id),\end{align*}
\begin{align*}
&=\left(\begin{array}{ccc}e^{-2itg_+(k,\xi)}&0\\\\f(k)&
e^{2itg_+(k,\xi)}\end{array}\right), & k\in(ic,id),\end{align*}
\begin{align*}
&=\left(\begin{array}{ccc}e^{-2itg_+(k,\xi)}&f(k)\\\\0&
e^{2itg_+(k,\xi)}\end{array}\right), & k\in(-ic,-id).
\end{align*}
\\\textbf{Step 2}\\
The function $g(k,\xi)$ and its imaginary part (see signature
table of the $\Im g(k,\xi)$ on the figure (\ref{Im g=0})) suggest a
choice of new contour $\Sigma_2$ for R-H problem. Let $L_\pm=\{k:
k=s\pm\mu(\xi), -\infty<s<\infty\}$ be the strait lines. Then
$\Sigma_2=\Db{R}\cup[-ic,ic]\cup L_+\cup L_-$. We use the
lower-upper factorization (\ref{flu1}) of the jump matrix on the
real axis  and apply the transformation

\begin{equation}
M^{(2)}(\xi,t,k)=M^{(1)}(\xi,t,k)G^{(2)}(\xi,t,k),\qquad
G^{(2)}(\xi,t,k)=\begin{cases}\left(\begin{array}{ccc}1&0\\\\-r(k)
e^{2itg(k,\xi)}&1
\end{array}\right),\quad k\in\Omega_1\\\\
\left(\begin{array}{ccc}1&\overline{r(\overline{k})}e^{-2itg(k)}\\\\0&1
\end{array}
\right),\quad k\in\Omega_2\\\\
I,\quad k\in\Omega_3\bigcup\Omega_4
\end{cases}
\end{equation}
where domains $\Omega_j, j=1,2,3,4$ are indicated on the figure
(\ref{Contourelliptic1}) and get the new R-H problem:
\begin{equation}
\quad M^{(2)}_-(\xi,t,k)=M^{(2)}_+(\xi,t,k)J^{(2)}(\xi,t,k),
\end{equation}
\[M^{(2)}(\xi,t,k)\rightarrow I,\quad k\rightarrow\infty\]
with the following jump matrices:
\begin{align*}
J^{(2)}(\xi,t,k)&=\begin{pmatrix} 1&0\\\\ -r(k) \ee^{2 \ii
tg(k)}&1
\end{pmatrix}, & k\in L_+, \end{align*}
\begin{align*} &=\begin{pmatrix}
1&\bar r(\bar k) \ee^{-2 \ii tg(k,\xi)}\\\\
0&1 \end{pmatrix}, &  k\in L_-,\\
&=\begin{pmatrix}
\ee^{ \ii t B_g(\xi)} & o \\\\
0& \ee^{\ii t B_g(\xi)}
\end{pmatrix},
& k\in (-id,id)\\  &=\begin{pmatrix} \ee^{-2 \ii tg_+(k,\xi)}
&0\\\\ f(k) & \ee^{2 \ii tg_+(k,\xi)}
\end{pmatrix},
& k\in (ic,id) , \\  &=\begin{pmatrix}
\ee^{-2 \ii tg_+(k,\xi)} & f( k) \\\\
0& \ee^{2 \ii tg_+(k,\xi)}
\end{pmatrix},
&  k\in (-ic,-id).
\end{align*}

We have just used the properties 1-5 of the function $g(k,\xi)$,
the jump relation $$f(k)=r_-(k)-r_+(k),\quad k\in(0,ic)$$ and
symmetry relation $$\quad r(k)=-\overline{r(\overline{k})},\quad
k\in
\left(\mathbb{C}\backslash(-ic,ic)\right)\\\bigcup\left((-ic,ic)_+
\bigcup(-ic,ic)_-\right).$$
\\\textbf{Step 3}\\
First we note that the function $f(k)$ has the following analytic
continuation from the  interval $(-ic,ic)$:
\begin{equation}f(k)=\widehat{f}_+(k),\quad k\in(-ic,ic),
\end{equation}
where
\begin{equation}\widehat{f}(k)=\displaystyle\frac{4}{\varkappa^2(k)-\displaystyle
\frac{1}
{\varkappa^2(k)}}=\frac{2i}{c}\sqrt{k^2+c^2}, \qquad
 \varkappa(k)=\sqrt[4]{\displaystyle\frac{k-ic}{k+ic}}.
\end{equation}
Then we  factorize the jump matrix $J^{(2)}$ on
$(ic,id)\bigcup(-ic,-\ii d)$ as follows:
\begin{equation}J^{(2)}(\xi,t,k)=
=F_+^{-\sigma_3}(k,\xi)\begin{pmatrix}1&0\\\\\displaystyle\frac
{e^{2itg_+(k,\xi)}}{\widehat{f}_+(k)F_+^2(k,\xi)}&1
\end{pmatrix}\begin{pmatrix}0&i\\i&0\end{pmatrix}
\label{f-ic-id}\begin{pmatrix}1&0\\\\\displaystyle\frac
{-e^{2itg_-(k,\xi)}}{\widehat{f}_-(k)F_-^2(k,\xi)}&1
\end{pmatrix}F_-^{\sigma_3}(k,\xi),
\end{equation}
for $k\in (ic,id)$ and
\begin{equation}J^{(2)}(\xi,t,k)=
F_+^{-\sigma_3}(k,\xi)\begin{pmatrix}1&\displaystyle\frac
{F_+^2(k,\xi)e^{-2itg_+(k,\xi)}}{\widehat{f}_+(k)}\\\\0&1
\end{pmatrix}\begin{pmatrix}0&i\\i&0 \end{pmatrix}
\label{ficid} \begin{pmatrix}1&\displaystyle\frac
{-F_-^2(k,\xi)e^{-2itg_-(k,\xi)}}{\widehat{f}_-(k)}\\\\0&1
\end{pmatrix}F_-^{\sigma_3}(k,\xi)\end{equation}
for $k\in (-ic,-id)$. Direct calculations show that it is
possible if
\begin{itemize}
\item $ F(k,\xi)$ is analytic outside the contour $ [ic,-ic]$
which is oriented from $ic$ to $-ic$.;
\item $ F(k,\xi)$ does not
vanish;
\item $ F(k,\xi)$ satisfies the jump relation:
\[
  F_-(k,\xi)  F_+(k,\xi)=\begin{cases}-if(k)=(a_-(k)a_+(k))^{-1},\quad
k\in(ic,id)\\\displaystyle\frac{i}{f(k)}=a_-(k)a_+(k),\quad
k\in(-ic,-id),\end{cases},
\]
and \[ F_-(k,\xi)=F_+(k,\xi) h(k),\quad k\in(-id,id),
\]
where $h(k)$ is some function, which has to be determined.
  \item
$ F(k,\xi)$ is bounded at infinity.
\end{itemize}

\begin{figure}[ht]
\begin{center}
\epsfig{width=100mm,figure=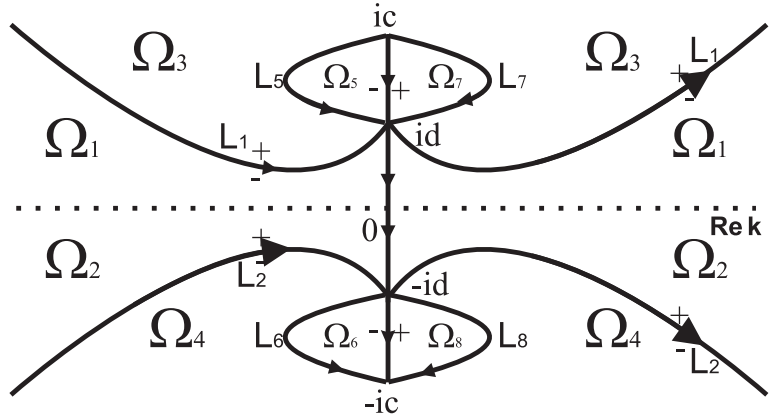}
\end{center}
\caption{The contour $\Sigma_3$ for the R-H problem $M^{(3)}(\xi,t,k)$}
\label{Contourelliptic2}
\end{figure}

To solve this RH problem let us put
\begin{equation}\label{F}F(k,\xi)=\begin{cases}\displaystyle\frac{1}{a(k)}
F_{aux}(k,\xi),
\quad k\in
\mathbb{C}_+\backslash[0,ic]\\\\a(k)F_{aux}(k,\xi),\quad k\in
\mathbb{C}_-\backslash[-ic,0]\end{cases},\end{equation} . Since
\[
  F_{aux-}(k,\xi)  F_{aux+}(k,\xi)=1,\quad k\in(ic,id)\bigcup(-ic,-id)
\]
\[ F_{aux+}(k,\xi)=F_{aux-}(k,\xi) a^2(k),\quad k\in\mathbb{R}
\]
We use the function $\mathrm{w}(k)=\sqrt{(k^2+c^2)(k^2+d^2)}$.Since
\[
\left[\frac{\log {F_{aux}(k,\xi)}}{\mathrm{w}(k)}\right]_+ -
\left[\frac{\log{F_{aux}(k,\xi)}}{\mathrm{w}(k)}\right]_-
=0, \qquad k\in   (ic,id)\bigcup(-ic,-id),
\]
and
\[
\left[\frac{\log {F_{aux}(k,\xi)}}{\mathrm{w}(k)}\right]_+ -
\left[\frac{\log{F_{aux}(k,\xi)}}{\mathrm{w}(k)}\right]_-
=\frac{\log{a^{2}(k)}}{\mathrm{w}(k)}, \qquad k\in\Db{R},
\]
we have that one of the functions which satisfy the last two equations is
\begin{equation}\label{Faux}
\widetilde{F}_{aux}(k,\xi)=\exp\left\{\frac{\mathrm{w}(k)}{2\pi\ii}
\int\limits_{\Db{R}} \frac{\log a^{2}(s)}{s-k}\frac{ds}{\mathrm{w}(s)}
\right\}.
\end{equation}
\[\widetilde{F}_{aux}(k,\xi)=e^{ik\Delta}\left(1+\ord\left(\displaystyle\frac{1}
{k}\right)
\right),k\rightarrow\infty\]
where\[\Delta=\displaystyle\frac{1}{2\pi}\displaystyle\int\limits_{-\infty}
^{\infty}
\frac{\log a^2(s)ds}{\mathrm{w}(s)}<0\]
But $\widetilde{F}_{aux}(k,\xi)$ has an essential singularity in infinity. To remove this singularity we introduce an Abelian integral of the second kind
$$\Omega(k)=d_0+\displaystyle\int\limits_{ic}^{\infty}\frac{(s^2+e_0)ds}
{\mathrm{w}(s)}$$
so that
$$\Omega(k)=k+\ord\left(\displaystyle\frac{1}{k}\right),\quad k\rightarrow\infty$$
$$\displaystyle\int\limits_{0}^{id}\frac{(s^2+e_0)ds}{\mathrm{w}(s)}=0$$

As
\begin{equation}
\begin{array}{ccc}
\Omega(k)-k\in \Db{R},\quad k\in \Db{R}\\
\Omega(k)-k\in i\Db{R},\quad k\in i\Db{R}
\end{array}
\end{equation}
then $d_0=0$ and $$\Omega(k)=\displaystyle\int\limits_{ic}^{\infty}\frac{(s^2+e_0)ds}
{\mathrm{w}(s)}$$

Then \[\Omega_++\Omega_-=0\quad \textrm{on}\quad(ic,id)\bigcup(-ic,-id)\]
\[\Omega_--\Omega_+=B_{\Omega}<0\quad \textrm{on}\quad (id,-id)\]

Define
\begin{equation}
F_{aux}(k,\xi)=\exp\left\{\displaystyle\frac{\mathrm{w}(k)}{\pi
i}\int\limits_{-\infty}^{\infty}\displaystyle\frac{\ln
a(s)ds}{(s-k)\mathrm{w}(s)}-i\Delta(\xi)\Omega(k,\xi),\right\}\end{equation}

Then $F(k,\xi)$ defined by (\ref{F}) has the following additional properties:
\begin{itemize}
  \item $\lim\limits_{k\rightarrow\infty}F(k,\xi)=1,\quad k\rightarrow\infty$
  \item $F_-(k,\xi)=F_+(k,\xi)\displaystyle e^{-iB_{\Omega}(\xi)\Delta(\xi)},\quad k\in(-id,id),$
\end{itemize}
By using the factorizations (\ref{ficid}), (\ref{f-ic-id}) we get
the following RH problem:
\begin{equation}
M^{(3)}(\xi,t,k)=M^{(2)}(\xi,t,k)G^{(3)}(\xi,t,k),\quad
M^{(3)}_-(\xi,t,k)=M^{(3)}_+(\xi,t,k)J^{(3)}(\xi,t,k),
\end{equation}
\[M^{(3)}(\xi,t,k)\rightarrow I,\quad k\rightarrow\infty\]
where
\begin{equation}
G^{(3)}(\xi,t,k)=F^{-\sigma_3}(k,\xi)\left(\begin{array}{ccc}1&
\displaystyle
\frac{F^2(k,\xi)e^{-2itg(k,\xi)}}{\widehat{f}(k)}\\\\0&1
\end{array}
\right),\quad
k\in\Omega_5\bigcup\Omega_7
\end{equation}
\begin{equation}
=F^{-\sigma_3}(k,\xi)\left(\begin{array}{ccc}1&0
\\\\\displaystyle\frac{e^{2itg(k,\xi)}}{\widehat{f}(k)F^2(k,\xi)}&1\end{array}
\right),
\quad k\in\Omega_6\bigcup\Omega_8
\end{equation}
\begin{equation}
=F^{-\sigma_3}(k,\xi),\quad
k\notin\Omega_5\bigcup \Omega_6\bigcup\Omega_7\bigcup\Omega_8,
\end{equation}
and
\begin{equation}
J^{(3)}(\xi,t,k)=\left(\begin{array}{ccc}1&\displaystyle\frac{F^2(k,
\xi)
e^{-2itg(k,\xi)}}{\widehat{f}(k)}\\\\0&1\end{array}\right),\quad
k\in L_7,\qquad
=\left(\begin{array}{ccc}1&\displaystyle\frac{-F^2(k,\xi)
e^{-2itg(k,\xi)}}{\widehat{f}(k)}\\\\0&1\end{array}\right),\quad k\in L_5
\end{equation}
\begin{equation}
=\left(\begin{array}{ccc}1&0\\\\\displaystyle\frac{e^{2itg(k,\xi)}}{\widehat{f}
(k)F^2(k,\xi)}
&1\end{array}\right),\quad k\in L_8,\qquad
=\left(\begin{array}{ccc}1&0\\\\\displaystyle\frac{e^{2itg(k,\xi)}}{\widehat{f}
(k)F^2(k,\xi)}
&1\end{array}\right),\quad k\in L_6
\end{equation}
\begin{equation}
=\left(\begin{array}{ccc}1&0\\\\\displaystyle\frac{-r(k)e^{2itg(k,\xi)}}{F^2(k,
\xi)}&1
\end{array}\right),\quad k\in L_1\qquad
=\left(\begin{array}{ccc}1&\displaystyle\frac{-\overline{r(\overline{k})}F^2(k,
\xi)}
{e^{2itg(k,\xi)}}\\\\0&1
\end{array}\right),\quad k\in L_2
\end{equation}
\begin{equation}
=e^{(itB_g(\xi)+iB_{\Omega}(\xi)\Delta(\xi))\sigma_3},\quad
k\in(-id,id)\qquad
=\left(\begin{array}{ccc}0&i\\i&0\end{array}\right),\quad
k\in(ic,id)\bigcup(-id,-ic)
\end{equation}
\\\textbf{Step 4}\\Now we introduce a model problem $M^{(mod)}_-=M^{(mod)}_+J^{(mod)},\quad M^{(mod)}\rightarrow I\quad\textrm{as}\quad k\rightarrow\infty,\quad$ where
\begin{equation}
J^{(mod)}(\xi,t,k)=\begin{cases}
e^{(itB_g(\xi)+iB_{\Omega}(\xi)\Delta(\xi))\sigma_3},\quad k\in(-id,id)\\\\
\left(\begin{array}{ccc}0&i\\i&0\end{array}\right),\quad
k\in(ic,id)\bigcup(-id,-ic)
\end{cases}
\end{equation}
To solve it we need a notion of the Riemann surface $X$, which is
induced by $$\mathrm{w}^2(k)=(k^2+d^2)(k^2+d^2),$$ with cuts along
$(ic,id)$ and $(-id,-ic)$. On the first sheet of this surface
$\mathrm{w}(0)>0$. We introduce the a-cycle and the b-cycle as it
is shown at the picture:\\The basic of the holomorphic
differential forms is given by the differential form
\[\omega=2\pi
i\displaystyle\frac{\displaystyle\frac{dk}{\mathrm{w}(k)}}{\displaystyle
\int\limits_{a}\displaystyle\frac{dk}{\mathrm{w}(k)}}.\]Then
\[\displaystyle
\int\limits_{a}\displaystyle\frac{dk}{\mathrm{w}(k)}=2\pi i\]
\[\tau=\tau(\xi):=\displaystyle
\int\limits_{b}\displaystyle\frac{dk}{\mathrm{w}(k)}<0\] We
introduce the Poincare theta function:
\[\Theta(z)=\Theta(z,\tau(\xi))=\sum \limits_{m=-\infty}^{\infty}
\exp\left\{\displaystyle\frac{1}{2}\tau(\xi)m^2+zm\right\}
\]
It has the following property:
\begin{equation}\Theta(z+2\pi in+\tau(\xi) l,\tau(\xi))=\Theta(z,\tau(\xi))\exp\left\{-\displaystyle\frac{1}{2}\tau(\xi) l^2-zl
\right\},\quad n\in\mathbb{N},l\in\mathbb{N}\end{equation} Then
we introduce the Abel mapping on the surface $X$:
\[A:X\rightarrow\mathbb{C}/\left(2\pi i\mathbb{Z}+\tau(\xi)\mathbb{Z}
\right)\qquad
A(P)=\int\limits_{ic}^{P}\omega\] Now we introduce the functions
$\varphi(.,\xi),\psi(.,\xi):\{\textrm{the first sheet of the
X}\}\rightarrow \mathbb{C}$
\begin{equation}\label{phipsi}\begin{array}{l}\varphi_j(k,\xi)=\displaystyle
\frac{\Theta(A(k)-A(D_j)-K-
itB_g(\xi)-iB_{\Omega}\Delta(\xi))}
{\Theta(A(k)-A(D_j)-K)}\\\\
\psi_j(k,\xi)=\displaystyle\frac{\Theta(-A(k)-A(D_j)-K-itB_g(\xi)-iB_{\Omega}
\Delta(\xi))}
{\Theta(-A(k)-A(D_j)-K)}\end{array},\quad
j=\displaystyle\overline{1,2},\end{equation} where $D_1=(0,-cd)$
is on the second sheet and $D_1=(0,cd)$ is on the first sheet.
$A(D_1)=-A(D_2).$ $K$ is the Riemann constant of the surface $X$.
The integration contour in (\ref{phipsi}) is taken on the first
sheet and not to intersect the interval(-ic,ic). These functions
have the following properties:
\[\begin{array}{l}\varphi_+(k,\xi)=\psi_-(k,\xi)\\\\
\psi_+(k,\xi)=\varphi_-(k,\xi)\end{array},\quad k\in
(ic,id)\bigcup(-id,-ic)\]
\[\begin{array}{l}\varphi_-(k,\xi)=\varphi_+(k,\xi)e^{itB_g(\xi)+iB_{\Omega}
\Delta(\xi)}\\\\
\psi_-(k,\xi)=\psi_+(k,\xi)e^{-itB_g(\xi)-iB_{\Omega}\Delta(\xi)}\end{array},
\quad
k\in (id,-id)\] Define a function
\[\gamma(k)=\gamma(k,\xi)=\left(\displaystyle\frac{k-ic}{k-id}\right)^
{\displaystyle\frac{1}{4}}\left(\displaystyle\frac{k+id}{k+ic}\right)^
{\displaystyle\frac{1}{4}}\] Then the solution of the model
problem of the \textbf{4} can be produce as following:

\begin{equation}M^{(mod)}(\xi,t,k)=\left(\begin{array}{ccc}
M^{(mod)}_{11}(\xi,t,k)&M^{(mod)}_{12}(\xi,t,k)\\\\
M^{(mod)}_{21}(\xi,t,k)&M^{(mod)}_{22}(\xi,t,k)\end{array}
\right)\end{equation}

\begin{equation}\hfill M^{(mod)}(\xi,t,k)=
\end{equation}
\begin{equation}{\label{Mmodelliptic}}
\left(\begin{array}{ccc}\varphi_1^{-1}(\infty,\xi)&0\\0&\varphi_2^{-1}(\infty,
\xi)
\end{array}
\right)
\left(\begin{array}{ccc}\displaystyle
\frac{\gamma(k,\xi)+
\displaystyle\frac{1}{\gamma(k,\xi)}}{2}\varphi_1(k,\xi)&
\displaystyle\frac{\gamma(k,\xi)-
\displaystyle\frac{1}{\gamma(k,\xi)}}{2}\psi_1(k,\xi)\\\\
\displaystyle\frac{\gamma(k,\xi)-
\displaystyle\frac{1}{\gamma(k,\xi)}}{2}\varphi_2(k,\xi)&
\displaystyle\frac{\gamma(k,\xi)+
\displaystyle\frac{1}{\gamma(k,\xi)}}{2}\psi_2(k,\xi)
\end{array}\right)
\end{equation}

Let us note that
$$A(\infty)=\displaystyle\frac{\pi i}{2}$$
$$K=\displaystyle\frac{\pi i}{2}+\displaystyle\frac{\tau}{2}$$
$$A(0_\pm)=\mp\displaystyle\frac{\tau}{2}+\displaystyle\frac{\pi i}{2}$$
Taking it in (\ref{Mmodelliptic}) we have:
\begin{equation}{\label{Mmodelliptic_}}\hfill M^{(mod)}(\xi,t,k)=\left(\begin{array}{ccc}
\left(M^{(mod)}(\xi,t,k)\right)_{11}&\left(M^{(mod)}(\xi,t,k)\right)_{12}\\
\left(M^{(mod)}(\xi,t,k)\right)_{21}&\left(M^{(mod)}(\xi,t,k)\right)_{22}
\end{array}\right)
\end{equation}
and
\begin{equation}{\label{Mmodelellipticgamma11}}
\left(M^{(mod)}(\xi,t,k)\right)_{11}=
\displaystyle
\frac{\left(\gamma+
\displaystyle\frac{1}{\gamma}\right)}{2}
\displaystyle\frac{\Theta(A(P)-\displaystyle\frac{\pi i}{2}-itB_g-iB_{\Omega}\Delta)}
{\Theta(A(P)-\displaystyle\frac{\pi i}{2})}\frac{\Theta(0)}{\Theta(itB_g+B_{\Omega}\Delta)}
\end{equation}
\begin{equation}{\label{Mmodelellipticgamma12}}
\left(M^{(mod)}(\xi,t,k)\right)_{12}=
\displaystyle\frac{\left(\gamma-
\displaystyle\frac{1}{\gamma}\right)}{2}
\displaystyle\frac{\Theta(-A(P)-\displaystyle\frac{\pi i}{2}-itB_g-iB_{\Omega}\Delta)}
{\Theta(-A(P)-\displaystyle\frac{\pi i}{2})}\frac{\Theta(0)}{\Theta(itB_g+B_{\Omega}\Delta)}
\end{equation}
\begin{equation}{\label{Mmodelellipticgamma21}}
\left(M^{(mod)}(\xi,t,k)\right)_{21}=
\displaystyle
\frac{\left(\gamma-
\displaystyle\frac{1}{\gamma}\right)}{2}
\displaystyle\frac{\Theta(A(P)+\displaystyle\frac{\pi i}{2}-itB_g-iB_{\Omega}\Delta)}
{\Theta(A(P)+\displaystyle\frac{\pi i}{2})}\frac{\Theta(0)}{\Theta(itB_g+B_{\Omega}\Delta)}
\end{equation}
\begin{equation}{\label{Mmodelellipticgamma22}}
\left(M^{(mod)}(\xi,t,k)\right)_{22}=
\displaystyle
\frac{\left(\gamma+
\displaystyle\frac{1}{\gamma}\right)}{2}
\displaystyle\frac{\Theta(-A(P)+\displaystyle\frac{\pi i}{2}-itB_g-iB_{\Omega}\Delta)}
{\Theta(-A(P)+\displaystyle\frac{\pi i}{2})}\frac{\Theta(0)}{\Theta(itB_g+B_{\Omega}\Delta)}
\end{equation}

We can also solve the model problem by using a function
$$\lambda(k)=\sqrt[4]{\frac{k-ic}{k+ic}}\sqrt[4]{\frac{k-id}{k+id}}$$
Then we have that
\begin{equation}{\label{Mmodelliptic__}}\hfill M^{(mod)}(\xi,t,k)=\left(\begin{array}{ccc}
\left(M^{(mod)}(\xi,t,k)\right)_{11}&\left(M^{(mod)}(\xi,t,k)\right)_{12}\\
\left(M^{(mod)}(\xi,t,k)\right)_{21}&\left(M^{(mod)}(\xi,t,k)\right)_{22}
\end{array}\right)\end{equation}

\begin{equation}{\label{Mmodelellam11}}\left(M^{(mod)}(\xi,t,k)\right)_{11}=\displaystyle
\frac{\left(\lambda+
\displaystyle\frac{1}{\lambda}\right)}{2}
\displaystyle\frac{\Theta(A(P)-\displaystyle\frac{\pi i}{2}-itB_g-iB_{\Omega}\Delta)}
{\Theta(A(P)+\displaystyle\frac{\pi i}{2})}\frac{\Theta(\pi i)}{\Theta(itB_g+B_{\Omega}\Delta)}\end{equation}
\begin{equation}{\label{Mmodelellam12}}
\left(M^{(mod)}(\xi,t,k)\right)_{12}=\displaystyle\frac{\left(\lambda-
\displaystyle\frac{1}{\lambda}\right)}{2}
\displaystyle\frac{\Theta(-A(P)-\displaystyle\frac{\pi i}{2}-itB_g-iB_{\Omega}\Delta)}
{\Theta(-A(P)+\displaystyle\frac{\pi i}{2})}\frac{\Theta(\pi i)}{\Theta(itB_g+B_{\Omega}\Delta)}\end{equation}
\begin{equation}{\label{Mmodelellam21}}
\left(M^{(mod)}(\xi,t,k)\right)_{21}=\displaystyle
\frac{\left(\lambda-
\displaystyle\frac{1}{\lambda}\right)}{2}
\displaystyle\frac{\Theta(A(P)+\displaystyle\frac{\pi i}{2}-itB_g-iB_{\Omega}\Delta)}
{\Theta(A(P)-\displaystyle\frac{\pi i}{2})}\frac{\Theta(\pi i)}{\Theta(itB_g+B_{\Omega}\Delta)}\end{equation}
\begin{equation}{\label{Mmodelellam22}}
\left(M^{(mod)}(\xi,t,k)\right)_{22}=\displaystyle
\frac{\left(\lambda+
\displaystyle\frac{1}{\lambda}\right)}{2}
\displaystyle\frac{\Theta(-A(P)+\displaystyle\frac{\pi i}{2}-itB_g-iB_{\Omega}\Delta)}
{\Theta(-A(P)-\displaystyle\frac{\pi i}{2})}\frac{\Theta(\pi i)}{\Theta(itB_g+B_{\Omega}\Delta)}
\end{equation}

Comparing (\ref{Mmodelellipticgamma11}),(\ref{Mmodelellipticgamma12}),(\ref{Mmodelellipticgamma21}),(\ref{Mmodelellipticgamma22}) with (\ref{Mmodelellam11}),(\ref{Mmodelellam12}),(\ref{Mmodelellam21}),(\ref{Mmodelellam22}) respectively we take that
\begin{equation}{\label{Thetaidentities1}}\displaystyle\frac{A(P)-\displaystyle\frac{\pi i}{2}}{A(P)+\displaystyle\frac{\pi i}{2}}=\sqrt{\displaystyle\frac{c+d}{c-d}}\displaystyle\frac{\gamma(k)+
\displaystyle\frac{1}{\gamma(k)}}
{\lambda(k)+\displaystyle\frac{1}{\lambda(k)}}=\sqrt{\displaystyle\frac{c-d}{c+d}}
\displaystyle\frac{\lambda(k)-\displaystyle\frac{1}{\lambda(k)}}
{\gamma(k)-\displaystyle\frac{1}{\gamma(k)}}\end{equation}
\begin{equation}{\label{Thetaidentities2}}\displaystyle\frac{\Theta(0)}{\Theta(\pi i)}=\sqrt{\frac{c+d}{c-d}}\end{equation}

Then
\[q_{mod}(x,t):=\lim\limits_{k\rightarrow\infty}2ik\left(M^{(mod)}\left(\frac{x}
{12t},t,k\right)-I\right)_{21}=\lim\limits_{k\rightarrow\infty}2ik\left(
M^{(mod)}\left(\frac{x}
{12t},t,k\right)-I\right)_{12}=\]
\[=(c+d)\displaystyle\frac{\Theta(\pi i+itB_g(\xi)+iB_{\Omega}(\xi)\Delta(\xi))}
{\Theta(0)}\displaystyle\frac{\Theta(\pi i)}{\Theta(itB_g+iB_{\Omega}\Delta)}=\]
\begin{equation}{\label{qmod}}=\sqrt{c^2-d^2}\displaystyle\frac{\Theta(\pi i+itB_g(\xi)+iB_{\Omega}(\xi)
\Delta(\xi),\tau)}
{\Theta(itB_g(\xi)+iB_{\Omega}(\xi)\Delta(\xi),\tau)}=
\sqrt{c^2-d^2}\displaystyle\frac{\Theta(\pi i+iU,\tau)}
{\Theta(iU,\tau)}\end{equation}
where $U=t B_g+B_{\Omega}\Delta$.

$q_{mod}$ can be expressed in terms of Jacobi elliptic functions:

\begin{equation}{\label{qmodJacobi}}
q_{mod}(x,t,k)=(c+d)\textrm{dn}\left(K(m)\left(\displaystyle\frac{U}{\pi}+1\right)
|m\right)=\displaystyle\frac{c-d}{\textrm{dn}\left(K(m)\displaystyle\frac{U}
{\pi}|m\right)}
\end{equation}
where
\[\tau=\displaystyle\frac{-2\pi K(1-m)}{K(m)},\quad m\in(0,1)\]
\begin{equation}
K(m)=\displaystyle\int\limits_{0}^{\frac{\pi}{2}}\frac{d\theta}{(1-m \sin^2(\theta))^{\frac{1}{2}}}
\end{equation}

\subsubsection{\large Asimptotics solitons}
As $\xi\rightarrow -\displaystyle\frac{c^2}{2}$ then $d\rightarrow 0$ and $\tau\rightarrow-\infty$. Then $q_{mod}(x,t)\rightarrow c$ as $\xi\rightarrow -\displaystyle\frac{c^2}{2}$.

If we direct $\xi$ to $\displaystyle\frac{c^2}{3}$ then $\tau$ tend to 0 and theta-functions in (\ref{qmod}) are slowly-convergent. But we can use the Poisson summation formula and rewrite the formula (\ref{qmod}) in rapidly convergent
theta-functions.
\begin{equation}
\Theta(z,\tau)=\Theta\left(\displaystyle\frac{2\pi i z}{\tau},\displaystyle\frac{4 \pi^2}{\tau}\right)\sqrt{\displaystyle\frac{2\pi}{-\tau}}
\left(\exp{\displaystyle\frac{-z^2}{2\tau}}\right)
\end{equation}
Then \begin{equation}
q_{mod}(x,t)=\sqrt{c^2-d^2}\exp{\left(\displaystyle\frac{-\tau^*}{8}+
\displaystyle\frac{\tau^*}{4}\left(\displaystyle\frac{U}{\pi}+1\right)\right)}
\displaystyle\frac{
\Theta\left(\displaystyle\frac{\tau^*}{2}\left(\displaystyle\frac{U}{\pi}+1\right),
\tau^*\right)}{\Theta\left(\displaystyle\frac{\tau^*}{2}\displaystyle\frac{U}
{\pi},\tau^*\right)}
\end{equation}
where $\tau^*=\displaystyle\frac{4\pi^2}{\tau}$

Now we have to expand $\tau^*$, $B_g$, $B_{\Omega}$, ${\Delta}$ in some series which are 
convergent when $\xi$ is tend to $\displaystyle\frac{c^2}{3}$.

\begin{equation}
\int\limits_{b}\displaystyle\frac{dk}{\textrm{w}(k)}=
2\int\limits_{d}^{c}\displaystyle\frac{dy}{\sqrt{(c^2-y^2)(y^2-d^2)}}=
\displaystyle\frac{1}{\pi}\left(1+\displaystyle\frac{1}{2}h+
\displaystyle\frac{5}{16}h^2+\ord{(h^3)}\right)
\end{equation}

Let us introduce the method of solving the inverse Jacobi problem (\ref{}).
Let X be a Riemann surface of genus g, $\{a_j\}_{j=1}^g$, $\{b_j\}_{j=1}^g$-homological basis of surface X, $\{\omega_j\}_{j=1}^g$ is the basis of homological differential of the surface X such that $\int\limits_{a_j}\omega_l=2\pi i\delta_{jl}$ and $B$-is matrix of b-periods of the basis of holomorphic differentials, $A:X\rightarrow 2\pi i\mathbb{Z}^g+B\mathbb{Z}^g$-is Abelian mapping.
Let $f$ be a meromorphic function on the Riemann surface X with poles $\{Q_j\}_{j=1}^
{m}$, $D=\{P_l\}_{l=1}^{g}$-is nonspecial divisor which doesn't contain any pole of $f$. We know that $Theta(A(P)-A(D)-K,\tau)$ is Abelian integral of the third kind with zeroes in points of divisor $D$.

Let us integrate a differential form $f(P)d\ln\Theta(A(P)-A(D)-K,\tau)$ along the
border of the fundamental polygon of the Riemann surface X.

Then we get
\begin{equation}
\sum_{l=1}^{g}f(P_j)=\displaystyle\frac{1}{2\pi i}\sum_{l=1}^{g}\int\limits_{a_l}
f(P)\omega_l-\sum_{j=1}^{l}res_{Q_j}\left(f(P)d\ln\Theta(A(P)-A(D)-K,\tau)
\right)\end{equation}

There are exist curves $\gamma_U$:
\begin{equation}
\displaystyle\frac{c(x-4c^2t)}{\log\displaystyle\frac{v}{8}}=
\displaystyle\frac{U}{2\pi}+\ord{\left(\displaystyle\frac{\log(-\log v)}{\log v}\right)\quad \mathrm{as}\quad v:=1-\displaystyle\frac{3\xi}{c}\rightarrow 0}
\end{equation}
such that if point $(x,t)$ lies between two lines $\gamma_{\pi(2n-\delta)}$ and
$\displaystyle\gamma_{\pi(2n+\delta)}$ then \[q_{mod}(x,t)=\ord\left(v^{1-\delta}\right)\quad \mathrm{as}\quad v\rightarrow 0\quad 0\leq \delta<1\]
and if point $(x,t)$ lies between two lines $\gamma_{\pi(2n+1-\delta)}$ and
$\displaystyle\gamma_{\pi(2n+1-\delta)}$ then \[q_{mod}(x,t)=\displaystyle\frac{2c}{\cosh(arg)}\left(1+\ord\left(v+v^{2(1-
\delta)}\right)\right),\quad 0\leq \delta<1\]
where \[arg=2c(x-4c^2t)+(2n+1)\log(32c^3t)+\displaystyle\frac{2c\left(x-4c^2t\right)}
{\log\left(\displaystyle\frac{v}{8}\right)}+2\pi ic\Delta\left(c\right)+\]\[+\left(2n+1\right)\log\left(
\displaystyle\frac{c\left(x-4c^2t\right)}{\log\left(\displaystyle\frac{v}{8e}\right)}
+\ord\left(\displaystyle
\frac{\log^2\left(-\log\left(v\right)\right)}{\log\left(v\right)}\right)\right),\quad
v\rightarrow 0\]

\subsection{\large Vanishing dispersive asymptotics ($x>4 c^2 t$) }

To study asymptotic behavior of the Riemann-Hilbert problem
$RH_{xt}$ in the region $x>\omega^2t$ we have used well-known
technics from \cite{DZ93}, \cite{DIZ93}, \cite{DIZ}. The large
time asymptotics of the solution in this region is defined by the
phase function
$\theta(k)=\DS\frac{1}{4}\left(\frac{1}{k}+\frac{k}{\xi^2}\right)$,
where $\xi^2=t/4x$. Indeed, the stationary points of the phase
function $\theta(k)$ are real and equal to $\pm\xi$. We have
\[
\Im\theta(k)=\D\frac{\vert k\vert^2 -\xi^2}{4\vert
k\vert^2\xi^2}\Im k
\]
Therefore $\Im\theta(k)>0$ ($\Im\theta(k)<0$) for $k$ lying in
the lower (upper) half-disk and out of the upper (lower)
half-disk defined by the circle $\vert k\vert^2= \xi^2$ (Figure
\ref{im5}). For $\xi^2<|E|^2=1/4\omega^2$ (that is, for
$x>\omega^2t$) and for $k\in\gamma\cup\bar\gamma$ the jump matrix
$J^{(1)}(x,t,k)$  tends to the identity matrix  as $t\to\infty$.
Hence the contour $\gamma\cup\bar\gamma$ does not contribute to
the main term of the asymptotics, which is defined by the
stationary points $\pm\xi$ and has the order $O(t^{-1/2})$. This
asymptotics of the solution was done in \cite{M}:

\begin{teor}\label{ZMR}
The solution of the IBV problem (\ref{srs})-(\ref{bc1}) for
$t\to\infty$ in the region $x>\omega^2t$ has a quasi-linear
dispersive character, i.e. it is described by the Zakharov -
Manakov type formulas:
   \begin{align*}
    q(x,t)=&2\sqrt{\frac{\xi^3\eta(\xi)}{t}}
     \exp\left\{2\ii\sqrt{xt} - \ii\eta(\xi)\log\sqrt{xt}+
     i\varphi(\xi) \right\}+\\
     &+2\sqrt{\frac{\xi^3\eta(-\xi)}{t}}
     \exp\left\{-2\ii\sqrt{xt} +\ii\eta(-\xi)\log\sqrt{xt}+
     i\varphi(-\xi) \right\}
    +o(t^{-1/2}), \qquad t\rightarrow \infty,
     \nonumber
  \end{align*}
where the functions $\eta(k)$  and  $\varphi(k)$ are given by the
equations
  \begin{align*}
     \eta(k)&= \frac{1}{2\pi}\log \Big(1-\rho^2(k)\Big),\qquad
     \xi^2=\frac{t}{4x},\\
    \varphi(k)&= \frac{\pi}{4}-3\eta(k)\log 2
     - \arg \rho(k) -
\arg {\Gamma}(-\ii\eta(k))+
\frac{1}{\pi}\int_{-\xi}^{\xi}\log|s-k|d\log[1-\rho^2(s)].
  \end{align*}
Here $\Gamma(-\ii\eta(k))$ is the Euler gamma-function, and
$\rho(k)=\DS\frac{\varkappa^2(k)-1}{\varkappa^2(k)+1}$.
  \end{teor}

\subsection{Zakharov\textendash Manakov region $x>4c^2t$}
We have the following chain of the transformations of the R-H
problem (\ref{RH}): First we use
\begin{equation}
M^{(1)}(\xi,t,k)=M(\xi,t,k)G^{(1)}(\xi,t,k),\quad
M^{(1)}_-(\xi,t,k)=M^{(1)}_+(\xi,t,k)J^{(1)}(\xi,t,k),
\end{equation}
\begin{equation}
G^{(1)}(\xi,t,k)=\left(\begin{array}{ccc}1&0\\\\-r(k)e^{2it\theta
(k,\xi)}&1
\end{array}\right),\quad k\in\Omega_1
\end{equation}
\begin{equation}
=\left(\begin{array}{ccc}1&\overline{r(\overline{k})}e^{-2it\theta(k)}\\\\0&1
\end{array}
\right),\quad k\in\Omega_2
\end{equation}
\begin{equation}
=I,\quad k\in\Omega_3\bigcup\Omega_4
\end{equation}

\begin{equation}
J^{(1)}(\xi,t,k)=
\left(\begin{array}{ccc}1&0\\\\-r(k)e^{2it\theta(k,\xi)}&1
\end{array}
\right),\quad k\in L_1\end{equation}\begin{equation}
=\left(\begin{array}{ccc}1&-\overline{r(\overline{k})}e^{-2it\theta(k,\xi)}\\\\0&1
\end{array}\right),\quad k\in L_2\end{equation}\begin{equation}
=I,\quad k\in(-ic,ic)
\end{equation}
Now we consider a model problem:
$M^{(mod)}_-(\xi,t,k)=M^{(mod)}_+(\xi,t,k)J^{(mod)}(\xi,t,k)$,\[M^{(mod)}
(\xi,t,k)\rightarrow
I,\quad k\rightarrow\infty\] where
\begin{equation}J^{(mod)}(\xi,t,k)=I,\quad k \in (-ic,ic)
\end{equation}
$M^{(mod)}$ is trivially solvable: \[M^{(mod)}(\xi,t,k)=I\]
\[q_{mod}(x,t):=\lim\limits_{k\rightarrow\infty}2ik\left(M^{(mod)}
\left(\frac{x}{12t},t,k\right)-I\right)_{21}=0
\]
\begin{rem}
 In the paper \cite{KK} the asymptotic behavior of the
solution of the problem (\ref{srs}), (\ref{ic}), (\ref{bcgen})
was studied in a  neighborhood of the leading edge $x=\omega^2t$
in terms of asymptotic solitons. The problem of matching of the
elliptic wave with the asymptotic solitons and these solitons
with the vanishing  self-similar wave is much more complicated
and will be considered elsewhere.
\end{rem}


\end{document}